\documentclass[prd,twocolumn,amsmath,amssymb,floatfix,superscriptaddress,nofootinbib]{revtex4-1}
\usepackage{bm}
\usepackage{amsmath}
\usepackage{epsfig}
\usepackage{color}
\usepackage{natbib}
\usepackage{textcase}
\usepackage{graphicx}
\usepackage{ifthen}
\usepackage{graphicx}
\usepackage[utf8]{inputenc}
\usepackage{amssymb}
\usepackage{latexsym}
\usepackage{epstopdf}
\DeclareGraphicsExtensions{.ps, .png}
\usepackage{dcolumn}
\usepackage{footnote}
\usepackage{tabularx,ragged2e,booktabs}
\usepackage[normalem]{ulem}
\usepackage{float}
\restylefloat{table}

\newcommand{\refsec}[1]{section~\ref{sec:#1}}

\def\lsim{\mathrel{\raise.3ex\hbox{$$<$$\kern-.75em\lower1ex\hbox{$\sim$}}}}
\def\gsim{\mathrel{\raise.3ex\hbox{$$>$$\kern-.75em\lower1ex\hbox{$\sim$}}}}

\newcommand{\beq}{\begin{equation}}
\newcommand{\eeq}{\end{equation}}

\newcommand{\bea}{\begin{eqnarray}}
\newcommand{\eea}{\end{eqnarray}}

\newcommand{\muKarcmin}{\mu\mathrm{K\, arcmin}}

\definecolor{darkgreen}{cmyk}{0.85,0.2,1.00,0.2}
\definecolor{purple}{cmyk}{0.5,1.0,0,0}

\definecolor{ultramarine}{rgb}{0.07, 0.04, 0.56}
\definecolor{cadmiumgreen}{rgb}{0.0, 0.42, 0.24}
\definecolor{indigo(dye)}{rgb}{0.0, 0.25, 0.42}
\usepackage[linktocpage=true]{hyperref}
\hypersetup{
colorlinks=true,
citecolor=ultramarine,
linkcolor=cadmiumgreen,
urlcolor=indigo(dye),
pdfauthor={},
pdftitle={},
pdfsubject={}
}

\newcommand{\meas}{\frac{d^2\bm{l}}{(2\pi)^2}}

\newcommand{\temp}{\text{temp}}

\newcommand{\We}{\mathcal{W}^{E}}
\newcommand{\Wp}{\mathcal{W}^{\phi}}
\newcommand{\bl}{\bm{l}}
\newcommand\prim[1]{^{%
		\ifcase#1 \or\prime\or\prime\prime\or\prime\prime\prime\else\mathrm{\romannumeral #1}\fi}}

\newcommand{\lcut}{l_{\rm cut}}
\newcommand{\lmin}{l_{\rm min}}
\newcommand{\lmax}{l_{\rm max}}

\newcommand{\vecL}{\mathbf L}

\newcommand{\vecx}{\mathbf{x} }

\newcommand{\obs}{\textrm{obs}}

\newcommand{\lb}{\left [}
\newcommand{\rb}{\right ]}

\newcommand{\Cov}{\textrm{Cov}}

\newcommand{\deflect}{{\boldsymbol{\alpha}}}
\newcommand{\deflecti}{{\boldsymbol{\alpha}^{-1}}}
\newcommand{\normdeflect}{\alpha}

\newcommand{\Dop}{D} 
\newcommand{\Dopt}{D^\dagger}
\newcommand{\St}{X} 
\newcommand{\Stdat}{X^{\rm obs}} 
\newcommand{\Stdatdag}{X^{\rm{obs},\dagger}} 

\newcommand{\X}{X} 
\newcommand{\bg}{\boldsymbol{g}}

\newcommand{\lensit}{\textsc{LensIt}}
\newcommand{\n}{\boldsymbol{n}}

\begin{document}

\title{CMB Delensing with Neural Network Based Lensing Reconstruction in the Presence of Primordial Tensor Perturbations}

\author{Chen Heinrich}
\email{chenhe@caltech.edu}
\affiliation{California Institute of Technology, Pasadena, CA 91125, USA}

\author{Trey Driskell}
\affiliation{Department of Physics $\&$ Astronomy, University of Southern California, Los Angeles, CA 90007, USA}

\author{Chris Heinrich}
\affiliation{Polycam Inc., San Francisco, CA 94104, USA}

\begin{abstract}
The next-generation CMB experiments are expected to constrain the tensor-to-scalar ratio $r$ with high precision. Delensing is an important process as the observed CMB $B$-mode polarization that contains the primordial tensor perturbation signal is dominated by a much larger contribution due to gravitational lensing. To do so successfully, it is useful to explore methods for lensing reconstruction beyond the traditional quadratic estimator (QE) (which will become suboptimal for the next-generation experiments), and the maximum a posterior estimator (which is still currently under development). In Caldeira et al. 2020~\cite{Caldeira:2018ojb}, the authors showed that a neural network (NN) method using the ResUNet architectrue performs better than the QE and slightly suboptimally compared to the iterative estimator in terms of the lensing reconstruction performance. In this work, we take one step further to evaluate the delensing performance of these estimators on maps with primordial tensor perturbations using a standard delensing pipeline, and show that the \emph{delensing} performance of the NN estimator is optimal, agreeing with that of a converged iterative estimator, when tested on a suite of simulations with $r = 0.01$ and $r = 0.001$ for $12.7^{\circ} \times 12.7^{\circ}$ maps at a CMB-Stage~4 like polarization noise level $1\, \mu \rm{K\, arcmin}$ and 1' beam. We found that for the purpose of delensing, it is necessary to train and evaluate the NN on a set of CMB maps with $l<l_{\rm{cut}}$ removed, in order to avoid spurious correlations on the scales of interest for the final delensed $B$-mode power spectrum $l<l_{\rm{cut}}$, similar to what was known previously for the QE and the iterative estimator. We also present various NN training techniques that can be extended for a simultaneous treatment of foregrounds and more complex instrumental effects where the modeling is more uncertain.

\end{abstract}


\maketitle

\section{Introduction}
\label{sec:intro}

Recent cosmic microwave background (CMB) measurements have shown that the Universe is consistent with a $\Lambda$CDM model (e.g.~\cite{Aghanim:2018eyx}), in which initial perturbations that developed into large-scale structures we see today were seeded during a process called inflation. These primordial perturbations could be of scalar or tensor type\footnote{vector types are shown to decay quickly.}, the latter is also known as gravitational waves. The scalar perturbations generate anisotropies in the CMB temperature and $E$-mode polarization, while the tensor perturbations generate both $E$ and $B$-mode polarization.

While past CMB experiments have put precise constraints on the energy content, evolution and initial conditions of the Universe by observing the temperature and $E$-mode polarization, the tensor-to-scalar ratio $r$ can only be determined by measuring the $B$-modes precisely. The current upper bound on $r$ comes from the BICEP-KECK experiment,
$r < 0.036$ (95\% confidence level)~\cite{BICEPKeck:2022mhb}, and it is one of the major goals of next-generation ground-based CMB experiments such as BICEP-KECK, Simons Observatory and CMB-Stage 4 (CMB-S4) to detect or constrain $r$ to a higher precision: The forecasted sensitivity is $\sigma(r) = 0.003$ for both the BICEP-KECK~\cite{BICEPKeck:2022mhb} and Simons Observatory~\cite{SimonsObservatory:2018koc}, while CMB-S4~\cite{CMB-S4:2020lpa} could reach a $\sigma(r) \sim 5-8\times10^{-4}$, capable of a greater than $5\sigma$ detection of $r > 0.003$ signal, or in the absence of a detection, putting an upper limit of $r<0.001$ at the 95\% confidence level.

One crucial aspect of a successful measurement of $r$ is the removal of the gravitational lensing signal, a process called delensing. Gravitational lensing of the CMB occurs when CMB photons are deflected by intervening gravitational potential as they travel from the last surface of scattering. This deflection causes additional $B$-mode signal in our observations that are much higher than the primordial tensor signal, so properly characterizing and removing this effect is indispensable for precise and unbiased measurements of $r$.

Many efforts have been underway for developing delensing techniques which seek to remove the lensing effects in the data. External delensing uses external data such as the cosmic infrared background (CIB) observations or deep galaxy surveys as a proxy for the lensing potential, and has been shown to be successful in delensing the SPTPol data with CIB data from Herschel ~\cite{Manzotti:2017net}. Internal delensing, on the other hand, leverages the reconstruction of the lensing potential using CMB observations themselves, and has been demonstrated on ACT~\cite{Han:2020qbp} Planck~\cite{Larsen:2016wpa,Carron:2017vfg,Aghanim:2018oex} and  POLARBEAR~\cite{Adachi:2019caa}. As the noise level of the next-generation CMB experiments decreases to $\Delta_P \lesssim 2\,\muKarcmin$, internal delensing performance is expected to become more efficient, exceeding that of the external delensing~\cite{Sherwin:2015baa}.

Most measurements to date use the quadratic estimator (QE) for internal reconstruction of the lensing potential \cite{Hu_2002, Okamoto:2003zw, PhysRevD.59.123507}. It has been shown however, that the QE will no longer be an optimal estimator once the polarization instrumental noise level drops below the lensing $B$-mode level, at about $\sim 5\, \muKarcmin$~\cite{PhysRevD.69.043005}.
We expect this to the be case for the CMB Stage-4 experiment, so finding more optimal methods for delensing will be crucial.

A few alternatives have been explored in recent years. The most well established being the maximum a posteriori (MAP) estimate of the lensing potential, which was first proposed in Refs.~\cite{PhysRevD.68.083002, PhysRevD.69.043005} and later received an improved implementation in Ref.~\cite{Carron:2017mqf} (with accompanying code $\lensit$\footnote{$\lensit$: \url{ https://github.com/carronj/LensIt}}) capable of dealing with realistic instrumental effects such as anisotropic noise, beams and sky cuts without significant approximations. This solution represents the optimal solution and is achieved through an iterative process: First apply the quadratic estimator to get an estimate of the lensing potential, delens the data with it; then apply the quadratic estimator and delens again; iterate until convergence~\cite{Smith_2012}.

Alternatively, the authors of Refs.~\cite{Anderes_2015, PhysRevD.100.023509, Millea:2020cpw, Millea:2020iuw} have also explored the Bayesian framework to estimate the lensing potential, directly maximizing the posterior by running MCMC chains and simultaneously constraining $r$ and the amplitude of the lensing potential $A_{\phi}$, yielding better uncertainties on $A_{\phi}$ than a QE pipeline. The first simultaneous estimation of parameters and lensing reconstruction was applied to the SPTpol data in Ref.~\cite{Millea:2020iuw}.

Beside the above advances, the technique of deep learning has seen tremendous success in industry and found many applications within cosmology as well\footnote{See a compilation as of 2022 at \url{https://github.com/georgestein/ml-in-cosmology}.}. The authors of  Ref.~\cite{Caldeira:2018ojb} have demonstrated successful CMB lensing reconstructions using a convolutional neural network with a ResUNet architecture, which when trained, can take as input observed CMB $Q$ and $U$ maps and output estimates of the lensing potential and delensed $E$ maps. They found that the lensing reconstruction performance from the NN exceeded that of the QE and approached a theoretical estimate for the iterative estimator.

In this work, we extend the work of Ref.~\cite{Caldeira:2018ojb} by assessing the delensing performance of the NN estimator by directly applying a standard delensing pipeline on a suite of simulated observed CMB maps including primordial tensor perturbations. We compare the delensing performance against the QE and the iterative estimator obtained using $\lensit$, and show that the neural network estimator performs optimally, yielding delensed $B$-mode power that agrees well with that of a converged iterative estimator for our idealized setups. We also find that the NN we trained yields better delensed $E$-mode reconstruction than previously found in Ref.~\cite{Caldeira:2018ojb}. We take note of various training techniques we used, especially those for generating a large sample of training data with varying $r$-values and instrument noise levels, which can be useful for training NNs when seeing a distribution of plausible models is helpful.


The paper is divided as follows. In \refsec{cosmology}, we describe the background on CMB lensing reconstruction and summarize current delensing techniques relevant to this work. In \refsec{deep_learning} we give a brief review of deep learning techniques, concentrating on the ResUNet architecture. In \refsec{setup}, we describe our NN setup, including the generation of data, and the particular NN implementation and training procedure we use. In \refsec{recon}, we report the NN performance on lensing reconstruction using CMB maps including tensor perturbations, and compare with the QE and iterative estimator, as well as previous results in literature.
In \refsec{delensing_results}, we run the standard delensing pipeline using the NN lensing estimator and compare its performance to the QE and iterative estimators for $r = 0.01$ and $r = 0.001$. Finally, we summarize and conclude in \refsec{conclusion}.

%
%
%
%

\section{CMB Delensing}
\label{sec:cosmology}

We begin in \refsec{QE} by summarizing the current lensing reconstruction technique, the quadratic estimator, which we use as a baseline when comparing with neural network performance. Then we introduce the iterative estimator in \refsec{iterative_estimator}, which maximizes the posterior of the lensing potential by iteratively delensing the maps and represents the theoretical best estimate. We describe in \ref{sec:delensing} the internal delensing techniques and outline the map-level delensing procedure that we use to make predictions for the delensing performance.

\subsection{Quadratic Estimator}
\label{sec:QE}

Since the two-point correlations of lensed CMB modes $X$ averaged over realizations of the CMB fields has the form
\beq
\langle X(\bm{l}) X(\bm{l'}) \rangle_{\rm CMB} \propto \phi(\bm{L}),
\eeq
 one can write down, in the flat sky approximation, the quadratic estimator -- the minimum-variance estimator of the lensing potential that are quadratic combinations of the CMB observables $\alpha = XY$, where $X, Y \in \{T, E, B\}$~\cite{Hu:2001kj}
    \bea
    \label{eqn:QE}
    \hat{\phi}_{\alpha}^{\rm QE}&&(\bm{L}) = A^{\alpha}_{L} \int \meas
    F_{\alpha}(\bm{l}, \bm{l'})
    X^{\obs}(\bm{l})Y^{\obs}(\bm{l'}), \notag \\
    \eea
where $\bm{L} = \bm{l} + \bm{l'}$.
Here the weights $F_{\alpha}(\bm{l}, \bm{l'})$ have been chosen to minimize the variance of the quadratic estimator, and
    \beq
    A^{\alpha}_L = \left[ \int \meas f_{\alpha}(\bm{l}, \bm{l'}) F_{\alpha}(\bm{l}, \bm{l'})\right]^{-1},
    \eeq
    the normalization is chosen such that the estimator is unbiased.

   While the quadratic estimator will no longer be the optimal estimator (as opposed to a maximum a posterior estimator) for low-noise levels in polarization below about $\sim 5\, \muKarcmin$, it is still the most competitive form of lensing reconstruction technique for current experiments' noise levels. In particular, the $EB$ estimator is expected to have the highest signal-to-noise out of all possible combinations of quadratic pairs $\alpha$.

   For the $EB$ estimator, we have
     \beq
    F_{EB}(\bm{l}, \bm{l'}) = \frac{f_{EB}(\bm{l}, \bm{l'})}{C_l^{EE, \rm obs} C_{l'}^{BB, \rm obs}},
    \eeq
    and
    \beq
    f_{EB}(\bm{l}, \bm{l'}) = C_l^{EE} (\bm{l} \cdot \bm{L}) \sin 2\left( \psi_{\bm{l}} - \psi_{\bm{l}'} \right),
    \label{eq:f_EB}
    \eeq
    where $\psi_{\bm{l}}$ is the polar angle of $\bm{l}$ with respect to the axes used to define the Stokes parameters. Here $E$ and $B$ are the beam-deconvolved observed fields,
    and $C^{XX, \rm obs}_l$
    is the fiducial observed power spectrum used to inverse-variance filter
    the field $X$
    \beq
    C_l^{XX, \rm obs} = C_l^{XX} + N_l^{XX},
    \eeq
    where $C_l^{XX}$ are the fiducial lensed power spectrum and
    \beq
    N_l^{XX} = \Delta_X^2 e^{l(l+1)\sigma^2 / (8\mathrm{ln}2)},
    \eeq
is the noise power spectrum. Here $\sigma$ is the FWHM of the beam, $\Delta_{X}$ is the detector noise in units of $\muKarcmin$ and $\Delta_P = \sqrt{2}\Delta_T$ for fully polarized detectors. These expressions assume no primordial $B$-mode, which would induce an additional term in $f_{EB}$ proportional to the primordial $B$-mode power spectrum $\tilde{C}_l^{BB}$. We ignore this contribution since it is negligible given current limits on $r$.

Given that the neural network reconstruction will make use of both the observed $Q$ and $U$ maps, to make fair comparisons, we will use in this paper the minimum-variance combination of all the polarization pairs,
\beq
\hat{\phi}^{\rm QE}(\bm{L}) = \sum_{\alpha} w_{\alpha}(L) \hat{\phi}_{\alpha}^{\rm QE}(\bm{L}),
\eeq
instead of the $EB$-only estimator used in Ref.~\cite{Caldeira:2018ojb}.
Here the weights on the individual estimators are
\beq
w_{\alpha} = \frac{\sum_{\beta}(\bm{N}^{-1})_{\alpha \beta}}{\sum_{\beta \gamma}(\bm{N}^{-1})_{\beta \gamma}},
\eeq
where $\bm{N}$ is given by
\bea
N_{\alpha\beta}(L) = && A_{\alpha}(L) A_{\beta}(L) \int \frac{d^2 \bm{l}_1}{(2\pi)^2} F_{\alpha}(\bm{l}, \bm{l'}) [ F_{\beta}(\bm{l}, \bm{l'})  \notag \\
 && \times C_{l}^{X_{\alpha}X_{\beta}} C_{l'}^{Y_{\alpha} Y_{\beta}}  + F_{\beta}(\bm{l'}, \bm{l}) C_{l}^{X_{\alpha} Y_{\beta}} C_{l'}^{Y_{\alpha} X_{\beta}} ], \notag \\
\eea
where $\alpha = X_{\alpha}Y_{\alpha}$ and $\beta = X_{\beta} Y_{\beta}$.

The reconstruction noise of the minimum-variance estimator is then
\beq
N_L = \frac{1}{\sum_{\alpha \beta} (\bm{N}^{-1})_{\alpha \beta}}.
\eeq
The quadratic estimator and its reconstruction noise is computed with weights $F_{\alpha}$ that use the lensed CMB power spectra, unless otherwise specified.

%

\subsection{Iterative Estimator}
\label{sec:iterative_estimator}

While the quadratic estimator has been the estimator of choice for previous generations of CMB experiments, the next-generation CMB experiments will reach a noise level ($\lesssim 5 \muKarcmin$) for which the quadratic estimator is no longer the minimum variance estimator~\cite{PhysRevD.67.043001}. In order to achieve optimal reconstruction, the maximum a posteriori estimator was first constructed in Ref.~\cite{PhysRevD.67.043001} under simplifying assumptions, and later extended to work for more realistic survey conditions in Ref.~\cite{Carron:2017mqf}. In this work, we follow Ref.~\cite{Carron:2017mqf} for its framework and make use of the accompanying public code $\lensit$. We briefly summarize the formalism here and refer the reader to Ref.~\cite{Carron:2017mqf} for more details.

The maximum a posteriori formalism formally solves for the maximum of the posterior probability density function of the lensing potential. It assumes Gaussian unlensed CMB, noise and deflection fields, while accounting for beam effects, inhomogeneous noise, and realistic masks without approximations. Intuitively, it delenses the data using a QE at first, then applies the QE on the resulting maps with appropriately modified weights, and repeats this procedure until convergence. Delensing using this iterative estimator was shown to improve the forecasted error on the tensor-to-scalar ratio by a factor of $\sim 2$ compared to the QE for the CMB Stage-4 experiment~\cite{Carron:2017mqf}.

Mathematically, the iterative estimator is computed as follows. Let the observed CMB be modeled as
\beq
\St^{\rm obs} = B \Dop \X + n,
\eeq
where $B$ is the linear response matrix encoding the beam and pixel window effects, $D$ is the lensing operator, and $n$ is the noise.
The covariance of the data in pixel space is therefore
\beq
\label{anisoCov}
\textrm{Cov}_\deflect \equiv \langle \Stdat \Stdatdag\rangle =  B \Dop C^{\rm unl} \Dopt B^\dagger + N,
\eeq
where $\deflect$ is the deflection field due to lensing, $C^{\rm unl}$ is the covariance of the unlensed CMB and $N$ is the noise covariance matrix assumed to be diagonal in pixel space.

The log-likelihood of the observed CMB map is then
\beq
\label{lik}
\begin{split}
\ln \mathcal{L}(\Stdat | \deflect) &= -\frac 12 \Stdat\cdot \Cov_\deflect^{-1}\Stdat - \frac 12 \det \Cov_\deflect.
\end{split}
\eeq
Using a Gaussian prior on the lensing potential $\phi$ which is expected to be nearly linear, the log posterior can be written as
\beq
\ln p (\phi | \Stdat) = \ln \mathcal{L}(\Stdat | \phi) - \frac 12 \sum_\vecL \frac{\phi_\vecL^2}{C_\vecL^{\phi\phi}},
\eeq
where we have $\deflect = \nabla \phi$ assuming a pure gradient lensing deflection.

The iteration scheme used to arrive at the maximum a posteriori solution for $\deflect$ is the Newton-Raphson scheme. The starting point is the Wiener-filtered quadratic estimator
\beq
\label{alpha_0}
\deflect_0(\vecL) =  \frac{C^{\phi \phi}_L}{C_L^{\phi\phi} + N^{\phi\phi,\rm len}_L} i\vecL \:\hat \phi^{\rm QE}(\vecL),
\eeq
where $N^{\phi\phi}_L$ is the Gaussian reconstruction noise for the quadratic estimator built from $Q$ and $U$ polarization maps. The subsequent update at each iteration follows
\beq
\label{Newton}
\deflect_{N + 1} = \deflect_N +  \lambda \: H_N \bg_N,
\eeq
where $\bg_N$ and $H_N$ are the gradient and the curvature matrix at the $N$-th step respectively. The rate parameter $\lambda$ is set to $\lambda = 1/2$, appropriate for CMB-S4-like configurations.

The gradient and the inverse curvature matrix are the main elements of the Newton-Raphson iteration scheme needed to find the maximum a posteriori solution. The total gradient of the posterior with respect to the deflection can be written as
\beq
g_a^{\rm tot} \equiv \frac{\delta \ln p(\deflect|\Stdat  )}{\delta \normdeflect_a(\n)}
= g^{\rm QD}_a - g^{\rm MF}_a + g^{\rm PR}_a,
\eeq
where we have a piece quadratic in the data (QD), a mean-field piece (MF) that comes from the determinant of the covariance in the likelihood, and a prior piece (PR) which is the most straightforward to evaluate.

The QD piece can be computed using
\beq
g^{\textrm{QD}}_a(\n)  = \lb V_\deflect \Stdat\rb^{i} (\n) \lb W^{a}_{\deflect} \:\Stdat\rb_{i} (\n),
\label{gQD}
\eeq
where
\beq
\label{Leg1}
W^{a}_{\deflect} \:\St^{\rm dat}(\vecx) = \Dop \nabla_a \X^{\rm WF}_{\deflect}(\vecx),
\eeq
and
\beq
\label{Leg2}
V_\deflect \St^{\rm dat} = B^\dagger N^{-1}\lb \St^{\rm dat}  - B \Dop \X^{\rm WF}_{\deflect}\rb.
\eeq
These are straight forward to evaluate once the Wiener-filtered data
\beq
\label{filter}
X^{\rm WF}_\deflect = \lb \left( C^{\rm unl} \right)^{-1} + \Dopt B^\dagger N^{-1} B \Dop \rb^{-1} \Dopt B^\dagger N^{-1}\Stdat
\eeq
is constructed, which involves the inversion of large matrices using the conjugate gradient descent method.

The $D^{\dagger}$ operator in the Wiener-filtering step involves computing the inverse deflection $\deflecti_{N + 1}(\n)$, which is solved iteratively using the Newton-Raphson scheme and converges after 3 iterations:
\beq
\begin{split}
&\deflecti_{N + 1}(\n)
= \deflecti_N(\n) \\ &- M_\deflect^{-1}(\n + \deflecti_N(\n)) \cdot (\deflecti_N(\n) + \deflect(\n + \deflecti_N(\n))),
\end{split}
\eeq
where the inverse deflection $\deflecti$ is defined such that it remaps the deflected points back to themselves
\beq
\vecx + \deflect(\vecx) +  \deflecti(\vecx + \deflect(\vecx)) \equiv \vecx,
\eeq
and where $M$ is the magnification matrix due to the change of coordinates induced by $\deflect$:
\beq
\lb M_\deflect \rb_{ab}(\vecx) = \delta_{ab}  + \frac{\partial \normdeflect_a}{\partial x_b}(\vecx).
\eeq

The mean-field piece may be evaluated using a large number of data simulations by observing that
\beq\label{MF}
\begin{split}
g^{\rm MF}_a(\n)  &= \frac 12 \frac{\delta \ln \det \Cov_\deflect}{\delta \normdeflect_a(\vecx)} =  \left\langle g_a^{\textrm{QD}}(\n) \right\rangle.
\end{split}
\eeq
where the average is over data realizations. The $\lensit$ code offers two options for calculating the mean-field: $\textsc{pertMF}$ or $\textsc{simMF}$. The latter uses a number of simulations to calculate $g_a^{\textrm{QD}}(\n)$ at each iteration step, and can become expensive quickly. We use the quicker solution $\textsc{pertMF}$ which is an approximation that works well for our setup.

Beside the gradient, the Newton-Raphson iteration scheme also requires the curvature of the likelihood
\beq \label{Curv}
\left[H^{-1}\right]^{ab}_{\vecL \vecL'} \equiv - \frac{\delta^2 \ln p(\Stdat | \deflect)}{\delta \normdeflect_\vecL^a \:\delta \normdeflect^{b,*}_{\vecL'}}.
\eeq
This is evaluated iteratively using the limited memory Broyden-Fletcher-Goldfarb-Shanno (L-BFGS) update~\cite{d70f72bb6d4749a5a8e29137299f32ed}
which is able to account for the non-Gaussian and realization-dependent properties of the likelihood. The starting inverse curvature is
\beq
\label{H_0}
H_L^0 = \left( \frac{1}{N_L^{\phi\phi, \rm unl}} +\frac{1}{C_L^{\phi\phi}} \right)^{-1}.
\eeq
where the first term is the likelihood curvature and second is the prior curvature.

%

\subsection{Delensing Techniques}
\label{sec:delensing}

Given the reconstructed potential, we can use it to predict an approximate lensing $B$-mode template. This template can then be subtracted from the observed $B$-mode map to obtain the delensed map
\beq
\label{eq:b_del}
    B^{\text{del}} = B^{\obs} -B^{\temp}.
\eeq
Because the delensed $B$-modes power spectrum
\bea
\label{eq:clbb_delensed}
    \langle B^{\text{del}}(\bl)B^{\text{del}}(\bl')\rangle  \equiv (2\pi)^2\delta_D(\bl+\bl') C_{l}^{BB,\text{delensed}}, \notag \\
\eea
is often reconstructed in order to measure $r$, one needs to be careful of the spurious correlations that can be introduced between the template and the observed $B$-mode. More specifically,
\bea
    && \langle B^{\text{del}}(\bl)B^{\text{del}}(\bl')
    \rangle
    =  \langle B^{\obs}(\bl) B^{\obs}(\bl')\rangle \notag \\
    &-&2\langle B^{\temp}(\bl) B^{\obs}(\bl')\rangle + \langle B^{\temp}(\bl) B^{\temp}(\bl')\rangle. \notag\\
\eea
Because the template $B$-modes are constructed using the lensing potential estimate which depends on the observed $E$ and $B$ fields, the template $B$-modes are correlated with the observed $B$-modes~\cite{2011arXiv1102.5729T}.

To eliminate this correlation, we can employ a cut in the CMB multipoles $l_{\rm cut}$ when performing the lensing potential reconstruction, so that on the range of interest of the delensed $B$-modes $l<l_{\rm cut}$, the observed $B$-modes are not used for the lensing reconstruction calculation~\cite{Lizancos:2020fio}. The observed $E$-modes in the range $l<l_{\rm cut}$ however can still be lensed to construct the $B$-mode template (step 2 below).

To ensure fair comparison, the same $\lcut$ on the observed CMB fields are imposed for evaluating all three estimators that we consider: The quadratic estimator $\hat{\phi}^{\rm QE}$, the iterative estimator $\hat{\phi}^{\rm iter}$ and the neural network estimator $\hat{\phi}^{\rm NN}$. We will see later that this also amounts to not training the NN with input CMB maps including $l<\lcut$, because the trained model may retain information about the observed CMB on these scales, even when evaluated on CMB maps with $l \geq \lcut$.

We use the following procedure for our delensing pipeline, given a set of simulated observed $Q^{\rm obs}$ and $U^{\rm obs}$ maps (including primordial tensor perturbations, lensing, and instrument noise) and corresponding lensing estimate $\hat{\phi}$ performed on $Q^{\rm obs}$ and $U^{\rm obs}$ with $l\geq\lcut$.

\begin{enumerate}

    \item Transform the observed total $Q^{\rm obs}$ and $U^{\rm obs}$ to $E^{\rm obs}$ and $B^{\rm obs}$ maps.

    \item Wiener filter the observed $E$ map using
    \beq
    \label{eq:wiener_filter_E}
        \We_l\equiv \frac{C_l^{EE}} {C_l^{EE} + N_l^{EE}},
    \eeq
    to get an estimate of the $E$-mode for constructing the lensed $B$-mode template: $E^{\mathrm{WF}}(\bm l) = \We_l E^{\rm obs}(\bm l)$.

    \item Set $B$-modes to zero, and transform this new set of $E =E^{\mathrm{WF}}$ and $B=0$ maps into $Q$ and $U$ maps. The lensing operation is performed using $\lensit$ by displacing pixels on the $QU$ map according to deflections implied by the reconstructed lensing potential $\hat{\phi}$.

    \item Transform the lensed $Q$ and $U$ maps back to $E$ and $B$ maps, and use this $B$-mode map as the lensing template $B^{\rm temp}$.

    \item Finally, use Eq.~\ref{eq:b_del} to obtain the delensed $B$-mode map.

\end{enumerate}

The QE and NN reconstructed lensing potentials are Wiener-filtered using
\beq
\label{eq:wiener_filter_phi}
    \Wp_L \equiv \frac{{C}_L^{\phi\phi}}{{C}_L^{\phi\phi}+N_L^{\phi\phi}}.
\eeq
In practice, we use for the QE $N_L^{\phi\phi} \approx N_{L}^{(0)}$, the disconnected four-point contribution, which dominates over other higher-order terms such as $N_L^{(1)}$~\cite{Cooray:2002py}. For the NN estimator, we use an approximate simulation-derived noise curve which will be defined later in Section~\ref{sec:recon} using Eq.~\ref{eq:nlkk_NN} where $\kappa(\bl)= \frac{1}{2} l (l + 1)\phi(\bl)$.
The iterative estimator does not need to be Wiener-filtered as it is already the optimal solution.

The predicted delensed $B$-mode power spectrum for each type of estimator is then calculated by using Eq.~\ref{eq:clbb_delensed} where the ensemble average is replaced by the average over a set of $n_{\rm sim}$ simulations.
We will use this predicted $C_l^{BB, \rm{delensed}}$ in section~\ref{sec:delensing_results} as a metric to compare the delensing performance between the QE, the iterative estimator and the NN estimator.

Let us now move to describe the background on deep learning needed for performing the neural network lensing reconstruction.

\begin{figure*}
    \centering
    \includegraphics[width=\textwidth, keepaspectratio]{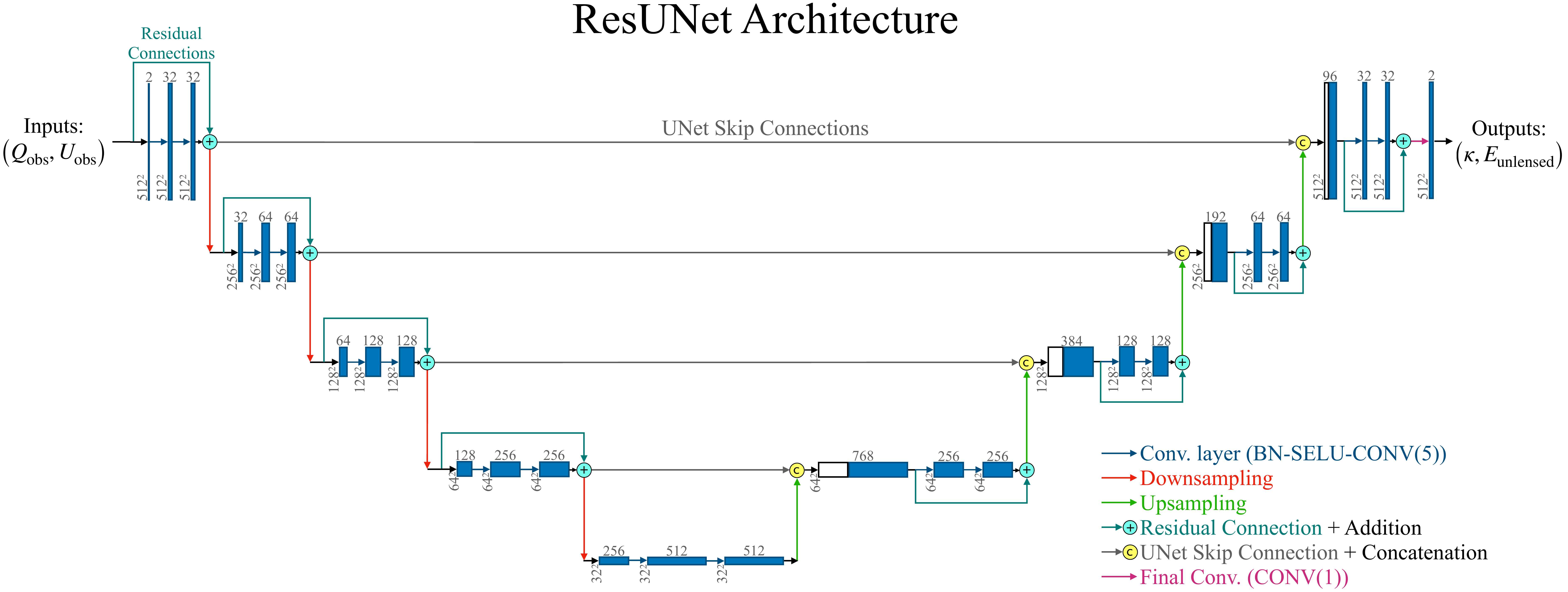}
    \caption{The particular implementation of the ResUNet architecture we used. The input are two channels of observed $Q$ and $U$ maps with $512\times512$ pixels, while the output are estimates of the $\kappa$ and $E_{\rm unlensed}$ maps. The ResUNet architecture combines both features of the UNet and the ResNet. The UNet is known for its U-shaped architecture with skip-connections (horizontal grey arrows) concatenating encoder-level (left-half) images with decoder-level (right-half) images. The ResNet is known for the residual skip-connections (cyan arrows) across the residual layer, optimizing the residual between the output and the input, instead of the output directly. Each residual layer here is composed of two convolutional layers (blue arrows), and is followed by a downsampling (red arrows) or upsampling (green arrows) operation, which halves or doubles the resolution respectively. The number of feature maps is increased at each residual level on the encoder (left) side, and decreased on the decoder (right) side. The data that flow through these operations are represented by the blue rectangles, where the number of feature maps are listed at the top, and where the resolution of individual maps are shown at the bottom. We use an initial number of feature maps $n_{\rm init} = 32$ at the first layer. See Section~\ref{sec:resunet} for more details on the ResUNet and Section~\ref{sec:neural_network_implementation} for our implementation.
    }
    \label{fig:resunet}
\end{figure*}

\begin{figure*}
    \centering
    \includegraphics[width=0.8\textwidth, keepaspectratio]{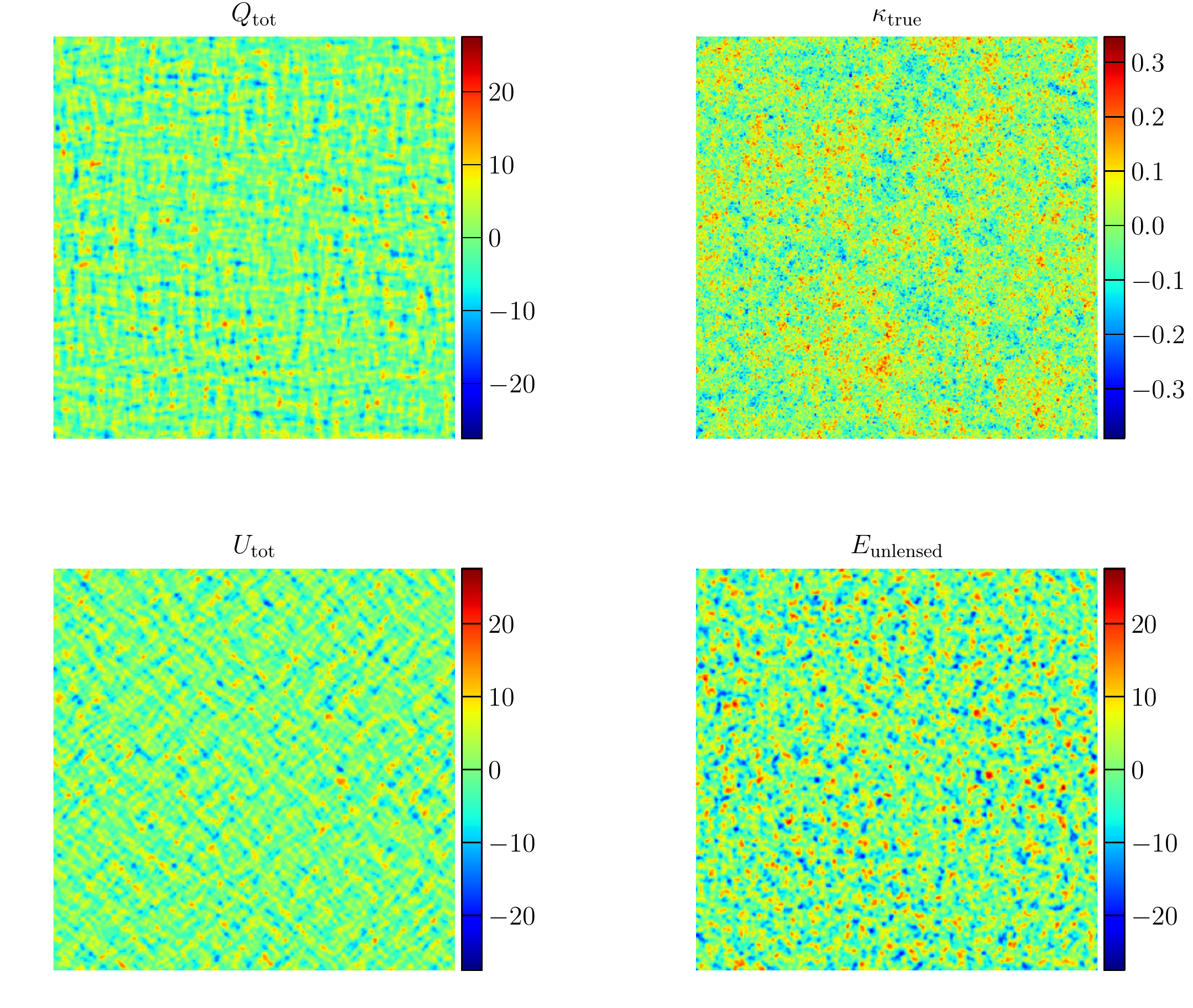}
    \caption{\emph{Left column}: The two input channels to the neural network -- the observed $Q$ and $U$ polarization maps with their multipoles cut out below $\lcut = 300$. \emph{Right column}: The two output channels of the neural network -- The true convergence $\kappa$ and unlensed $E$ polarization maps used as the target maps for training the neural network. Note that while we use a network with both output channels $\kappa$ and $E$ to reproduce (and improve upon) past work, the neural network used for obtaining the delensing results was trained using a single output channel $\kappa$.}
    \label{fig:maps}
\end{figure*}

\section{Deep Learning}
\label{sec:deep_learning}

In the last few years, deep learning has been applied to a variety of problems in cosmology. In the CMB field, it has been applied to infer foregrounds (e.g.~\cite{Farsian:2020adf, Petroff:2020fbf}), patchy reionization~\cite{Guzman:2021nfk}, lensing reconstruction~\cite{Caldeira:2018ojb, Li:2022kbd}, finding and applying Wiener filters~\cite{Munchmeyer:2019kng}, etc. In particular, its use for CMB lensing reconstruction was first shown by Ref.~\cite{Caldeira:2018ojb}.

Motivated by the success of Ref.~\cite{Caldeira:2018ojb} at performing lensing reconstruction, we leverage a version of a convolution neural network model~\cite{Lecun}, originally developed for the image-based learning tasks, for the purposes of B-mode template delensing.
In particular, we employ a ResUNet neural network as the deep learning model, and train the network using stochastic gradient descent, in which the weights in the network are trained to minimize a loss function that quantifies the distance between the NN output and the target maps. The result is a nonlinear transformation that, when applied on valid input maps, can provide an estimate of the desired target maps, which in our case is an estimate of the lensing potential map given input observed CMB maps.

In the following, we briefly summarize the major components of the ResUNet, and refer the readers to Ref.~\cite{Caldeira:2018ojb} for more details on ResUnet and to Ref.~\cite{goodfellow2016} for a more general introduction on deep learning.

\subsection{The ResUNet Architecture}
\label{sec:resunet}

The ResUNet is a type of Convolutional Neural Networks (CNN)~\cite{2018IGRSL..15..749Z}. CNNs are composed of a sequence of convolutional layers, typically followed by a set of fully connected layers at the end. Each convolutional layer performs a 2D convolution on its input images, with a kernel size specifying the receptive field of each output pixel. For example, a kernel size of 5 means that each output pixel receives information from an area of 5x5 pixels in the input image, so the connections between pixels in the input and output images in a single convolutional layer are local. Fully connected layers, on the other hand, connect all pixels in the input images to the output images.

Each convolutional layer is composed of a set of linear transformations specified by weights (also called neurons) that can be updated during the training process in order to minimize the loss function. The linear transformation is typically followed by an activation function (for we will choose the rectified linear unit (ReLU) \cite{inproceedings}),  which adds nonlinearity to the operation and a max-pooling operation which downsizes the images by taking the maximum value over, for example, a 2x2 field. The convolutional layer is a powerful setup that allows for arbitrary nonlinear transformations to be approximated given enough layers and neurons~\cite{2018arXiv180510769Z}.

The ResUNet was originally developed for road extraction from aerial images~\cite{2018IGRSL..15..749Z}, and combines the particular features of both a ResNet~\cite{2015arXiv151203385H} and a UNet (developed for biomedical image segmentation~\cite{2015arXiv150504597R}). See Fig.~\ref{fig:resunet} for an illustration of the specific implementation used in this paper. The UNet can be illustrated with an overall U shape: During the first portion of the network called the encoder, the dimension of the images decreases with each layer; then a bottleneck is reached, after which there is the decoder part of the network, with each layer leading to a larger image dimension, mirroring the encoder part. The UNet is known for the skip connections concatenating images at the encoder level with the images at the corresponding decoder level, combining feature information from the encoder level with spatial information from the decoder level.

The number of feature maps after the first layer and the depth of the encoder, decoder, and any other additional layers sets the total number of neurons in the network. The feature maps are generated from the input maps using the convolutional layers and, when properly trained, extract useful features that represent relevant information inside the maps. If the number of feature maps after the first layer is set to $n_{\rm init}$, then the first layer of the encoder transforms $n_{\rm in}$ input channels into $n_{\rm init}$ output channels, the second layer transforms $n_{\rm init}$ into $2 n_{\rm init}$ channels, the third, $2 n_{\rm init}$ to $4 n_{\rm init}$, etc.

For the decoder in a UNet, each residual block also receives as input the output of the corresponding layer in the encoder, so that the last layer in the decoder receives $2 n_{\rm init}$ channels from the previous decoder layer, and $n_{\rm init}$ layer from the first encoder layer, such that it has a total number of $2 n_{\rm init} + n_{\rm init}$ input channels, and outputs $n_{\rm init}$ channels. Similarly, the second to last layer in the decoder receives a total of $4 n_{\rm init} + 2n_{\rm init}$ channels and outputs $2 n_{\rm init}$ channels. The lasts convolutional layer in the decoder is usually followed by a series of fully connected layers, or convolutional layers, that transform $n_{\rm init}$ channels into the final number of desired output channels $n_{\rm out}$.

The ResNet, also called the residual network, is known for its residual block consisting of two or more convolutional layers, with residual skip connection connecting the beginning of the residual block to the end, allowing the input to the residual block to be passed directly to the end and be added to the output of the final convolutional layer, so that the neural network optimizes the residual of the output with respect to the input instead of the output directly. These residual skip connections make it easier to train very deep networks, i.e. networks with many layers, without running into the so-called vanishing gradient problem that can otherwise occur when doing back-propagation through very deep networks.

The ResUNet is therefore a UNet composed of residual blocks.

\subsection{The Training Process}

The training is done through stochastic gradient descent, by minimizing a user-defined loss function that is a function of the output of the NN and the given target images. Because of the limited memory of the GPU, it is generally not possible to process all the images at the same time, especially for large datasets. Instead, the loss function is computed for batches of images at a time. The batch size refers to the number of input images used for each batch; traditionally, we call it a training epoch when all the input images have been used once, which may contain many batches.

The ADAM optimizer is a commonly used optimizer, which is an adaptive variant of the original stochastic gradient descent algorithm~\cite{2014arXiv1412.6980K}. with a specified learning rate for updating the weights. One can also make use of a learning rate scheduler, which could gradually reduce the learning rate every few epochs, or only reduce the learning rate when a performance metric has reached a plateau. An example of the plateau scheduler would be to reduce the learning rate by half if the validation error has not improved after 3 epochs. Hyperparameters such as the batch size, the learning rate, and the scheduler choice are typically tuned to suit the problem at hand.

Additional techniques can be used to stabilize the training process. For example, batch normalization is used to normalize the output of each layer into a standard normal distribution across the batch, before feeding it into the activation function. This generally leads to faster and more stable training~\cite{2018arXiv180511604S}


\section{Neural Network Setup}
\label{sec:setup}

\subsection{Map Generation}
\label{sec:map_generation}

To generate simulations of the observed CMB $Q$ and $U$ maps, we use the software $\lensit$. We consider a CMB-S4 like experiment with $1\ \muKarcmin$ noise and 1 arcmin beam. We adopt the LensIt internal resolution parameters (LDres, HDres) = (9, 10), which makes maps with an area of about $12.7^{\circ} \times 12.7^{\circ}$, with $512 \times 512$ pixels. The lensing operation is actually performed on higher resolution maps with $1024 \times 1024$ pixels, while the observed maps are given at the lower resolution. Although the unlensed $E$ and $B$ maps are generated at higher resolution for the lensing operation, we save these, along with the lensing convergence $\kappa$, at the observed lower resolution in order to facilitate neural network training. Given this setup, the largest mode accessible is $l = 28$. We will plot results of Fourier transforms starting at the second nonzero multiple $l = 40$ in later sections.

To produce the observed $Q$ and $U$ maps, we first generate the unlensed $E$-mode maps from scalar perturbations (for which there are no $B$-modes) and the unlensed $B$-mode maps from tensor perturbations (for which the $E$-modes are of comparable amplitude to the $B$-modes, and so contribute negligibly to the total $E$-modes including from the scalar perturbations, for the small $r$ values we consider)~\cite{Hu:1997hv}. We only perform lensing operations on the scalar-induced $E$-mode maps to produce lensed $Q$ and $U$ maps which are then added to $Q$ and $U$ maps obtained from the unlensed tensor-induced $B$-mode maps. In principle, there is also a contribution from lensing the tensor $B$-mode maps, but we will neglect this effect since it is a much smaller second-order effect. Finally, we add to the total lensed $Q$ and $U$ maps the effects of beam and instrument noise to create the observed maps used for training the NN.

All maps are generated at a fixed fiducial $\Lambda$CDM cosmology consistent with Planck best-fit: Primordial power spectrum amplitude $A_s = 2.14 \times 10^{-9}$ and tilt  = $n_s = 0.968$, matter density $\Omega_c h^2 = 0.118$, baryon density $\Omega_b h^2 = 0.0223$, neutrino density $\Omega_\nu h^2 = 6.45\times 10^{-4}$ (1 massive neutrino), Hubble constant $H_0 = 67.9$ and reionization optical depth $\tau = 0.067$.

\subsection{Neural Network Implementation}
\label{sec:neural_network_implementation}

We use a particular implementation of the ResUNet architecture which is publicly available on Github\footnote {\url{https://github.com/galprz/brain-tumor-segmentation}} (see Fig.~\ref{fig:resunet} for a detailed illustration). We use the framework \texttt{pytorch}\footnote{\url{https://pytorch.org/}}~\cite{2019arXiv191201703P} for training. We choose an encoder and decoder with 4 residual blocks each; the encoder and decoder are separated by a double convolutional layer between them, and there is 1 additional convolutional layer at the end of the decoder.

For the particular implementation of the ResUNet blocks, we use the default settings in the public code. Each residual block consists of a double convolutional layer with a residual skip connection as a CONV(1)-BN operation between the input of the first layer and the output of the second one. Each convolutional layer is composed of the operations batch-normalization, SELU-type activation function and a 2D convolution with kernel size 5, which we abbreviate to BN-SELU-CONV(5). For the encoder, a downsampling step happens at the very end of the residual block (instead of using max-pooling, we use the default method provided by the code, which is a convolutional layer with CONV(3)-BN-SELU where the convolution operation has stride 2). For the decoder, an upsampling step followed by the UNet skip-connection happens before the entire residual block. We choose to use $n_{\rm init}$ = 32 feature maps output by the first residual layer.

The inputs are multi-dimensional arrays of shape $2 \times 512 \times 512$, describing a pair of observed CMB $Q$ and $U$ maps including primordial tensor perturbations, while the output channels are $1 \times 512 \times 512$, trained on target maps of $\kappa$ in our baseline setups, or $2 \times 512 \times 512$ if unlensed $E$ is included. Fig.~\ref{fig:maps} illustrates a set of sample input maps and target maps. Each of the input map channels shown are rescaled by the standard deviation of that channel over the whole training set before being fed into the network, to facilitate training by keeping the dynamic range of inputs close to order unity.

We perform data generation during training by randomly pairing pre-generated maps with only scalar perturbations, maps with only tensor perturbations at $r_0 = 0.1$ and noise maps at $1\ \muKarcmin$ scaling appropriately the tensor maps with $\sqrt{r/r_0}$ and the noise maps for the appropriate noise levels. Combining the data this way can in principle produce $19200^3$ combinations of training maps for a fixed $r$ and noise level. In reality, we do not use all the available combinations, as we train over a limited number of epochs (usually 40). During each epoch, the NN is trained over a set of 19200 randomly paired data.

In contrast to traditional training which iterates over the same dataset every epoch, our NN sees a different set of maps in each epoch by exploiting the additive property of our input data. This is called sampling without replacement. It is not an uncommon practice, though it is less well understood theoretically. The main motivation here is that we want the NN to see as many examples as possible over the course of the training. But we do not want to define one epoch as cycling through all possible combinations, which would mean a much slower convergence rate since the gradients are updated at the end of each epoch, and so is the validation loss which is needed to control the learning rate scheduling. Indeed, the NN converged after seeing only $40 \times 19200$ randomly drawn examples out of all $19200^3$ possible examples.

We explore training on a distribution of input $r$ as well as on random noise levels. We sample $r$ values from a log uniform distribution from $r_{\rm min}$ to $r_{\rm max}$ and perform validation during training on a set of maps with fixed $r_{\rm val}$. For example, our baseline results come from using $[r_{\rm min}, r_{\rm max}] = [0.0001, 0.1]$ and $r_{\rm val} = 0.01$. Similarly for the noise level, we train over input maps with noise levels uniformly distributed in the range [0, $2\Delta_{P, \rm val}$] where $\Delta_{P, \rm val} = 1\ \muKarcmin$ is the fixed noise level used for validation, so that the neural network can see examples of how the noise impacts the inputs.

We train the NN using the implementation readily available in \texttt{pytorch}. The loss function is chosen to be the L2 norm between the output and target maps averaged over the batch. We use a small batch size of 4, given the limited memory of our GPUs. We use the ADAM optimizer with an initial learning rate of 0.001 and use a plateau scheduler which decreases the learning rate by half if validation error has not improved after 3 epochs.

\section{Lensing Reconstruction Performance}
\label{sec:recon}

\begin{figure}
    \centering
    \includegraphics[width=0.47\textwidth, keepaspectratio]{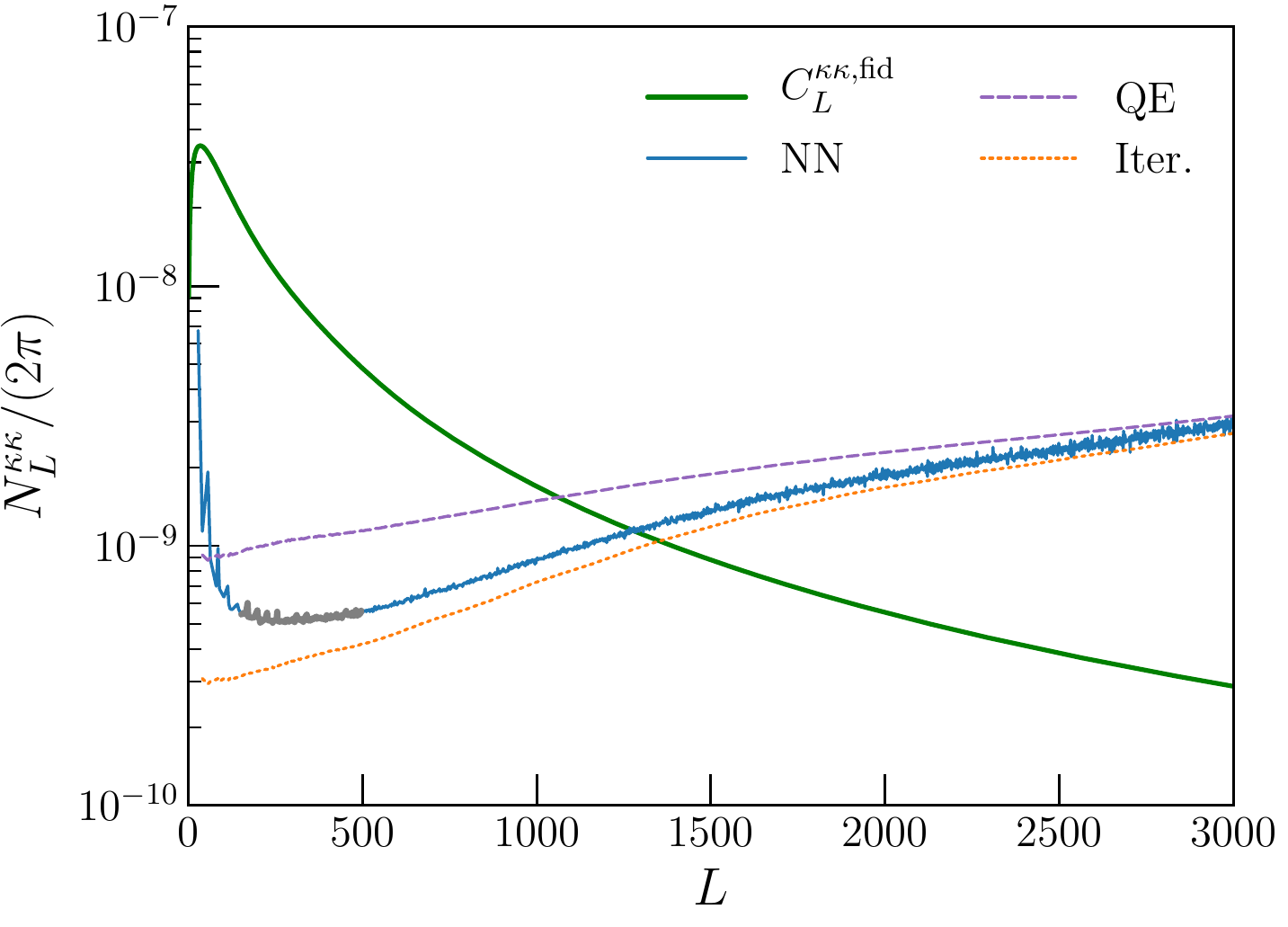}
    \caption{The reconstruction noise curve for various estimators: Quadratic estimator (purple dashed), iterative estimator (orange dotted), neural network (blue solid). The NN result is obtained by evaluating the trained model on the validation set with $n_{\rm sims} = 2400$ simulations at fixed $r = 0.01$ and fixed fiducial noise level. The quadratic and iterative estimator curves here are analytic approximations computed with no cut on the CMB multipoles; the weightings used in these estimators used $C_l^{BB, \rm lensed}$ with $r = 0$. The fiducial signal $C_L^{\kappa \kappa}$ (green solid) is shown here for reference. Note that the QE and iterative estimators curves are analytic, whereas the NN curve is simulation-based. The grey portion of the NN curve is used for averaging estimate in Fig.~\ref{fig:scaling}.}
    \label{fig:nlkk}
\end{figure}

\begin{figure}
\includegraphics[width=0.5\textwidth, keepaspectratio]{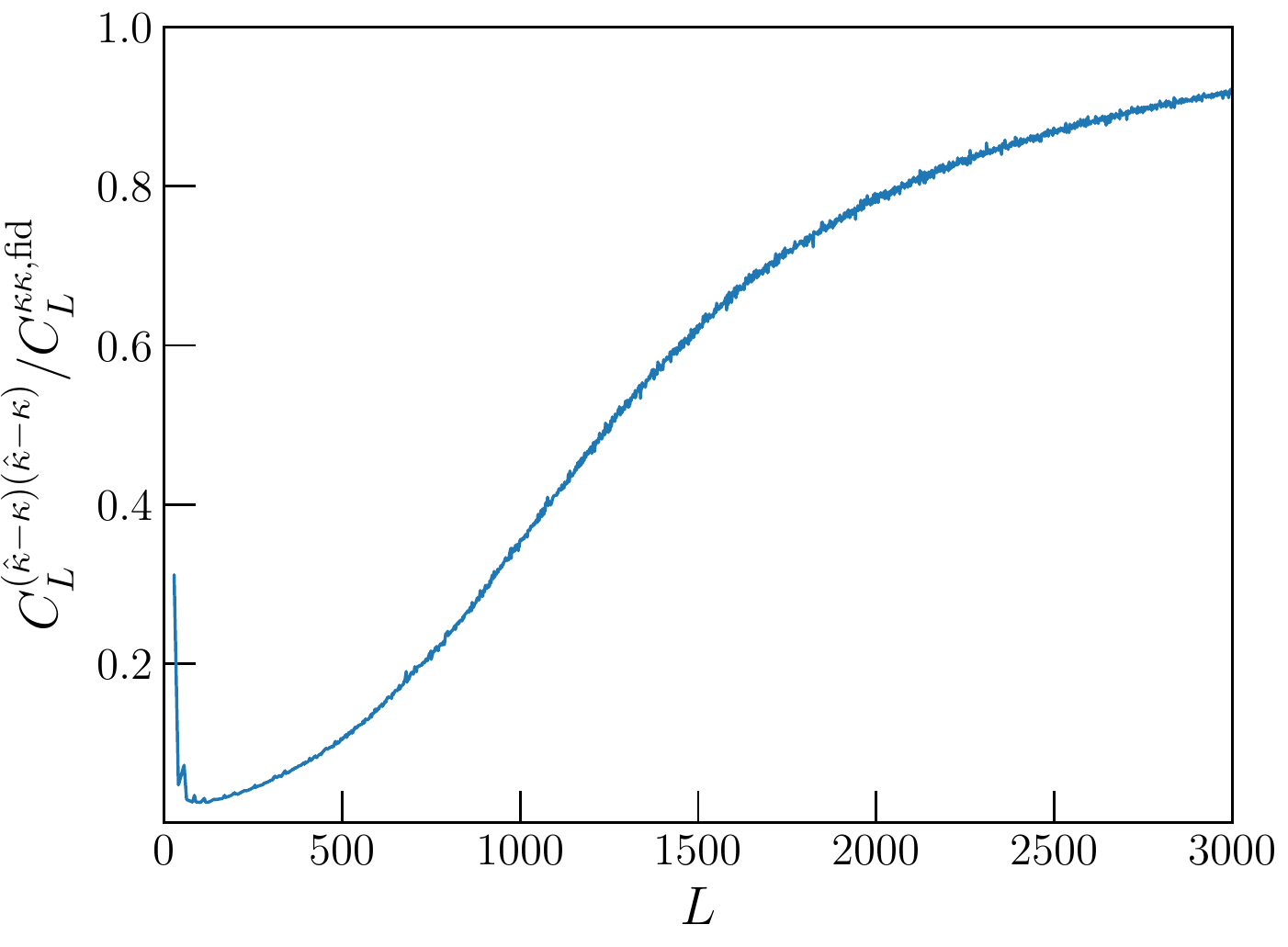}
\includegraphics[width=0.5\textwidth, keepaspectratio]{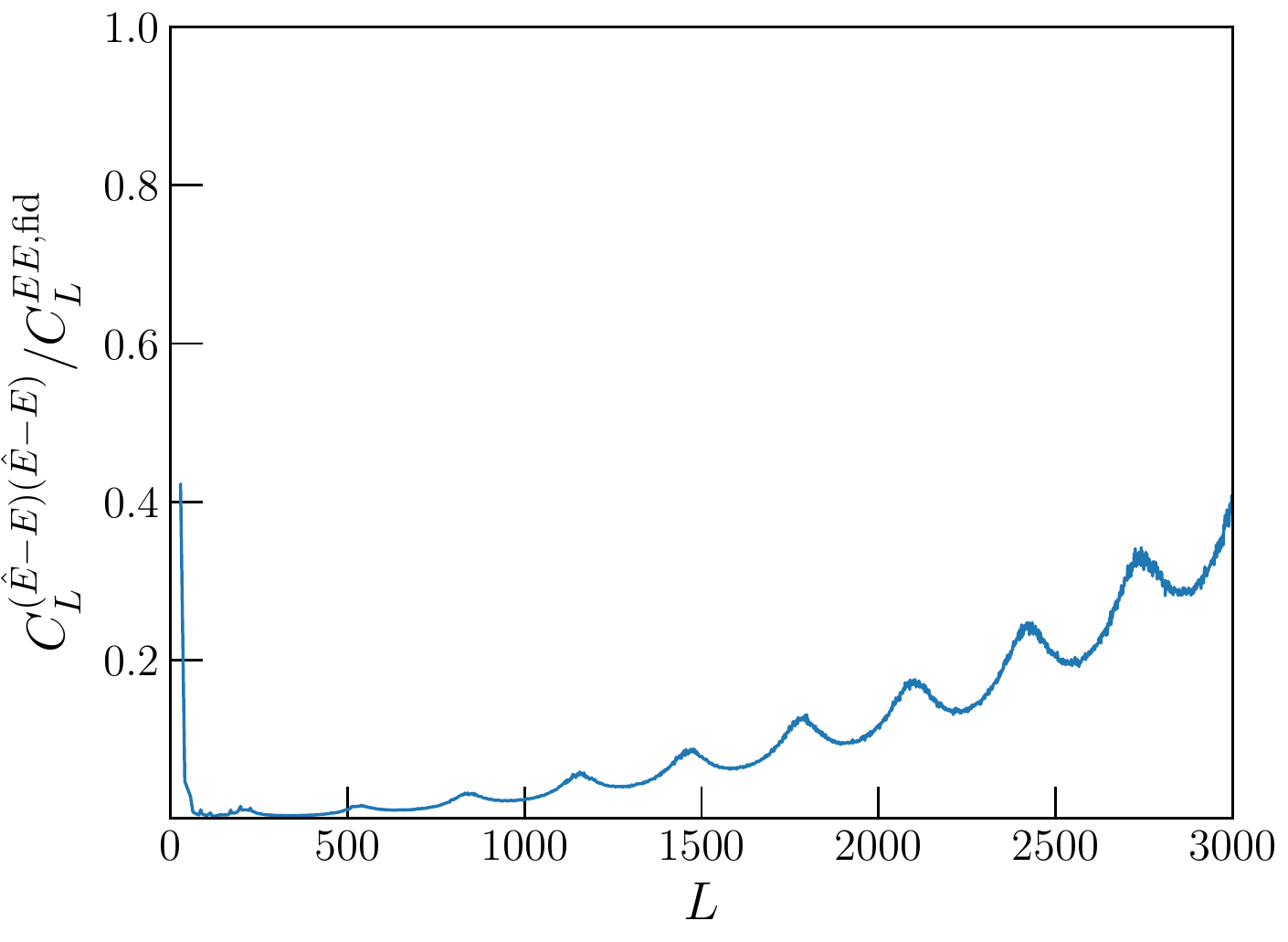}
\caption{The difference-map auto-spectrum $C_L^{(\hat{X} - X)(\hat{X} - X)}$ for the output channel $X$ (where the difference is taken between the pre-debiased neural network output $\hat{X}^{\rm out}$ and the true $X$), normalized to the fiducial signal $C_L^{XX, \rm fid}$. \emph{Top}: Lensing convergence channel $X = \kappa$. \emph{Bottom}: Unlensed $E$ channel $X = E_{\rm unlensed}$. The difference spectrum is obtained by averaging over the validation set with $n_{\rm sims}$ = 2400 simulations at fixed $r = 0.01$ and default noise level, using the neural network trained with two output channels $E$ and $\kappa$ and on CMB maps without $\lcut$ for literature comparison purposes. }
\label{fig:fractional}

\end{figure}

\begin{figure}
    \centering
    \includegraphics[width=0.45\textwidth, keepaspectratio]{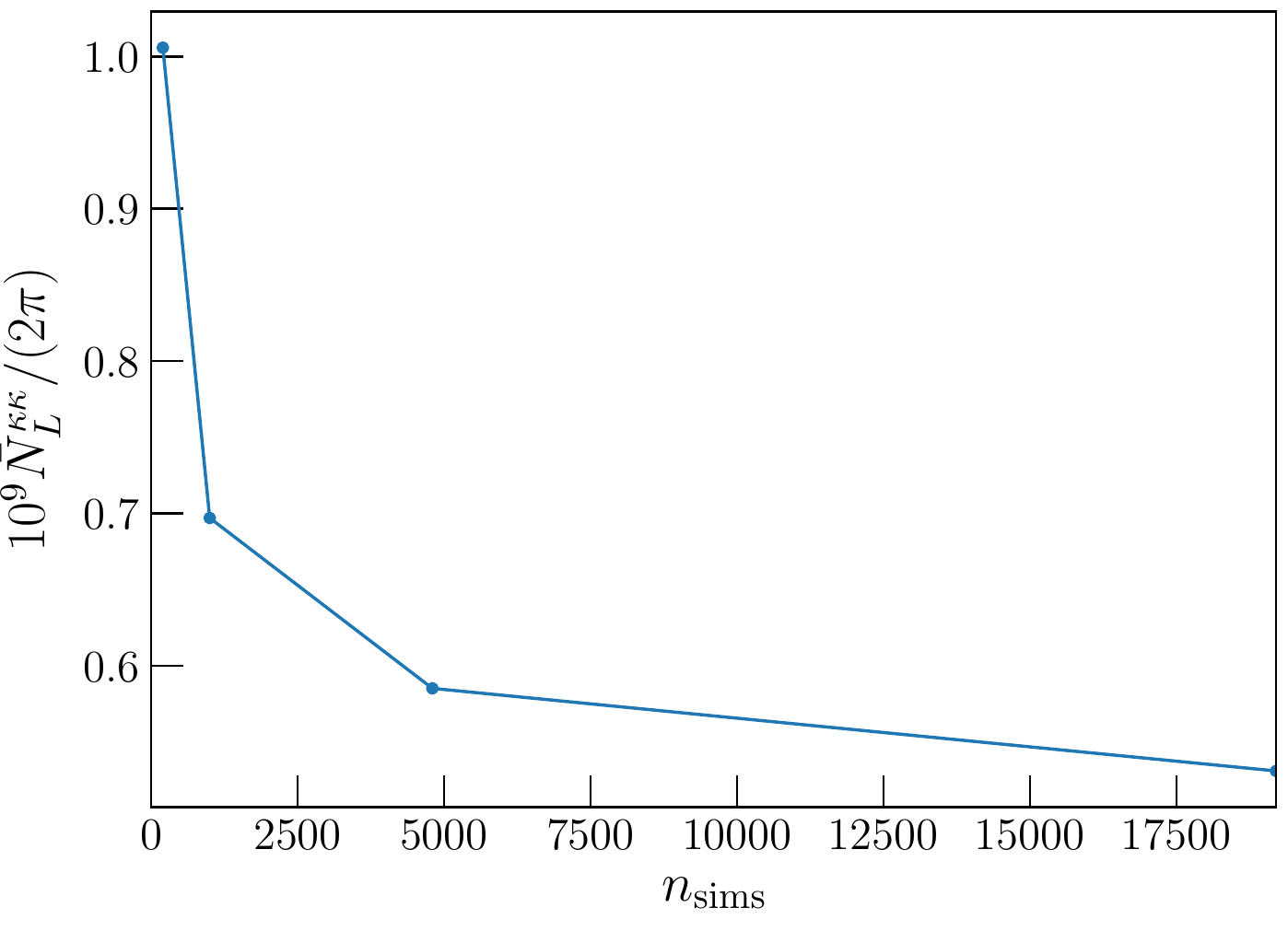}
    \caption{Approximate metric for the NN lensing reconstruction performance as a function of training data size $n_{\rm sim,\, train}$. We use $\bar{N}_L^{\kappa\kappa}$, the averaged NN reconstruction noise in the range $150 \leq L \leq 500$ as the approximate metric (see the corresponding grey portion in Fig.~\ref{fig:nlkk}). Beyond our fiducial choice of $n_{\rm sim}=19200$ (the rightmost point), we enter the regime of diminishing return, needing many more simulations (hence more disk space) for slightly better performance.}
    \label{fig:scaling}
\end{figure}

We now study the performance of our neural network on the task of lensing reconstruction. To facilitate comparison with previous literature, we will use, in this section only, the NN trained on two output channels $\kappa$ and $E_{\rm unlensed}$. Later in the delensing section, we will make use of the NN trained on a single output channel $\kappa$, so that the NN can focus on optimizing only the $\kappa$ performance.

To obtain the NN predictions for $\kappa$, we start by evaluating the trained NN model on a set of 2400 validation maps with fixed $r_{\rm val} = 0.01$. To analyze the output of the NN on this dataset, we follow Ref.~\cite{Caldeira:2018ojb} by first debiasing the direct output of the NN $\kappa_L^{\rm out}$ using
\beq
\hat{\kappa}_L^{\rm NN} = \frac{\langle \kappa_L^{\rm true} \kappa_L^{\rm true*} \rangle_{\rm val}}{\langle \kappa_L^{\rm out} \kappa_L^{\rm true *}\rangle_{\rm val}} \kappa_L^{\rm out},
\eeq
such that the cross-correlation with the true maps $\kappa_L^{\rm true}$ recovers
\beq
\langle \hat{\kappa}_L^{\rm NN} \kappa_L^{\rm true*} \rangle_{\rm val} = \langle \kappa_L^{\rm true} \kappa_L^{\rm true*} \rangle_{\rm val},
\eeq
and where the averages are over the validation set.

We define also the noise power spectrum from the NN estimator for a given $r_{\rm val}$ as
\beq
N_L^{\kappa\kappa} = \langle \hat{\kappa}_L^{\rm NN} \hat{\kappa}_L^{\rm NN*} \rangle_{\rm val} -
\langle \kappa_L^{\rm true} \kappa_L^{\rm true*} \rangle.
\label{eq:nlkk_NN}
\eeq
This can be directly compared to the analytic noise power spectrum for the quadratic estimator and iterative estimator, for which we choose to use lensed CMB spectra $C_l^{BB, \rm lensed}$ with $r = 0$ for the weighting used in the estimator. We have also verified that using $r = 0.01$ in the weighting of the quadratic estimator leads to negligible differences in the noise curve because of the low value of $r$.

In Fig.~\ref{fig:nlkk}, we show the noise power spectrum $N_L^{\kappa\kappa}$
from the NN estimate (blue solid), the quadratic estimator (purple dashed) and the iterative estimator (orange dotted), where the last two are evaluated with CMB $\lmax = 6000$. The fiducial signal $C_L^{\kappa\kappa}$ is shown in dark green for reference. The NN performance is obviously better than the QE and nears the iterative estimator line which is the theoretically optimal solution. Note that both the QE and iterative estimator noise curves are computed analytically using $\lensit$, while the NN one is derived from simulations.

\begin{figure*}
    \centering
    \includegraphics[width=0.8\textwidth, keepaspectratio]{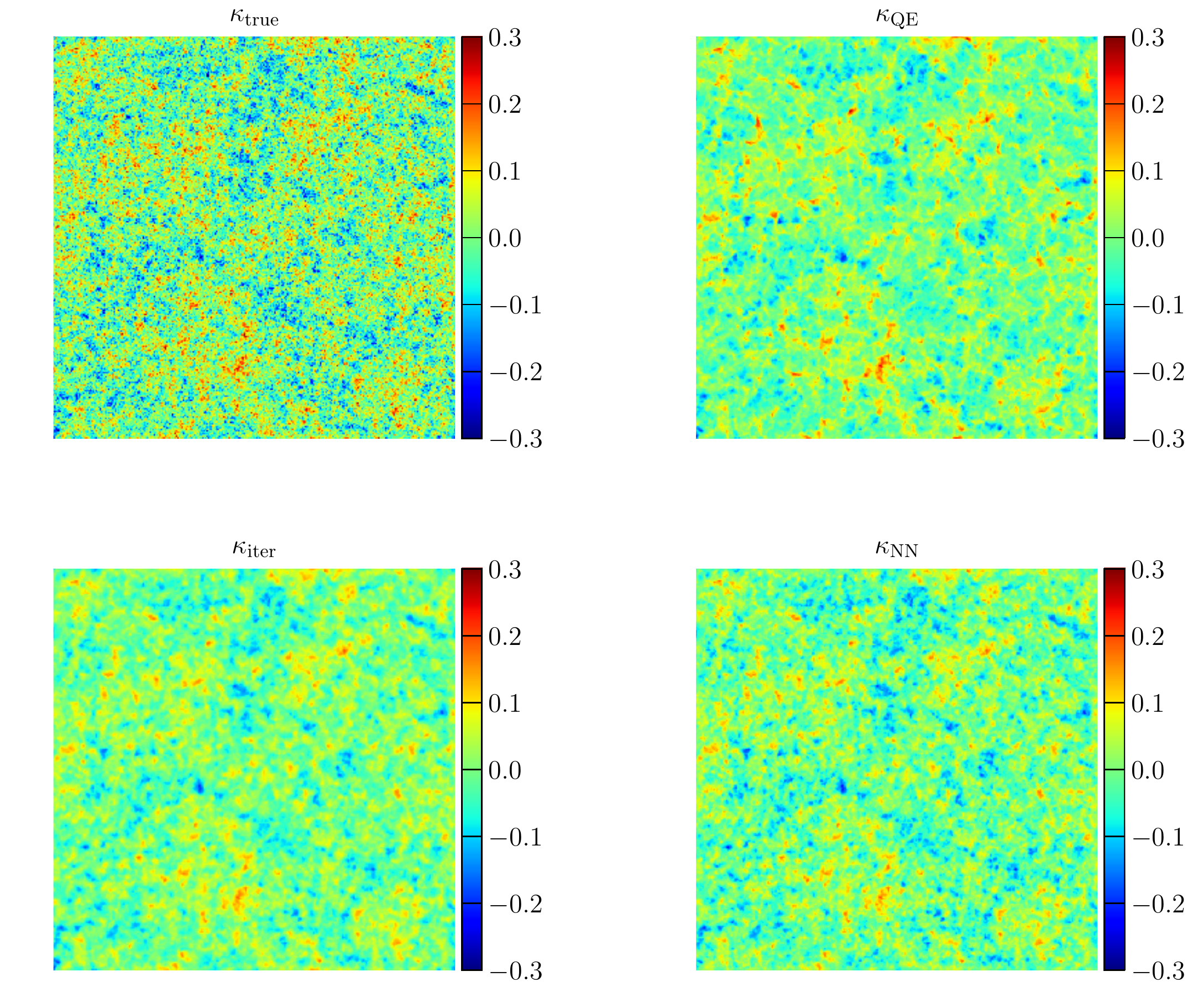}
    \caption{Convergence maps $\kappa$ as used in the delensing pipeline. \emph{Top left}: The true $\kappa$ map used for training the neural network. \emph{Top right}: The Wiener-filtered quadratic estimator applied to the observed $Q$ and $U$ maps in Fig.~\ref{fig:maps}. \emph{Bottom left}: The iterative estimator applied to the same $Q$ and $U$ maps after $n_{\rm iter} = 10$ iterations, which is converged. \emph{Bottom right}: The Wiener-filtered neural network estimate obtained by evaluating the trained network on the same $Q$ and $U$ maps. Features in the NN result are similar to those in the iterative estimator, and contain more details than those in the QE, which corresponds to the $n_{\rm iter} = 0$ iteration of the iterative estimator. All three estimators lose information on the smaller scales that are originally in the true $\kappa$ map due to the cosmic variance and other sources of error contributing to the reconstruction noise.}
    \label{fig:kappa_maps}
\end{figure*}

Furthermore, in order to compare most fairly with the NN results (which are evaluated using the $Q$ and $U$ maps), we use the minimum-variance (MV) weighted combination of all polarization estimators for the QE and iterative estimator instead of approximating using the $EB$ estimator alone as done in Ref.~\cite{Caldeira:2018ojb} (which could sometimes underestimate QE performance by a factor of 2 for certain ranges of $L$ at $\Delta_P \sim 1-2\, \muKarcmin$~\cite{Hu_2002}).
Finally, we checked that evaluating the NN on a validation set with $r = 0.001$ gives a similar reconstruction noise curve as that shown here, so the plot holds qualitatively for $r = 0.001$ as well.

We show in Fig.~\ref{fig:fractional} the ratio of the difference spectrum to the fiducial power spectrum for $\kappa$ and $E$ in the top and bottom panels respectively. The difference maps are obtained by taking the difference between the NN output $X^{\rm out}$ (before debiasing) with the true map $X^{\rm true}$. For $\kappa$, the performance is best at low-$L$, where the ratio reaches half at about $L = 1250$ and becomes close to 1 at $L = 3000$ where there is much less meaningful information recovered. The performance is very similar to that seen in the DeepCMB study in Ref.~\cite{Caldeira:2018ojb} (c.f. the $1\, \muKarcmin$ curve in their Fig.~8).

We found however, a better $E$ performance on the large scales: The ratio reaches about 0.4 at $L = 3000$ here, whereas in the DeepCMB study it reaches about 0.9 at $L = 3000$, and attains 0.4 much more quickly at about $L = 1750$ where we have only 0.1. A similar case was also observed by the authors of Ref.~\cite{Guzman:2021nfk}, who reconstructed the patchy reionization optical depth $\tau(\hat{n})$ in addition to $\kappa$ and $E$ using a modified ResUnet architecture: They found that their $\kappa$ reconstruction performance matched that of Ref.~\cite{Caldeira:2018ojb}, but also had better $E$-mode reconstruction. Our results are roughly consistent with theirs for the $1\, \muKarcmin$ noise level. We also note that in both the $\kappa$ and $E$ cases, we trained the network on random $r$ values, and evaluated the plotted metric on a set of validation maps with a nonzero tensor contribution $r = 0.01$. It seems therefore that the inclusion of tensor perturbations at this level does not degrade the NN performance.

One natural question to ask is whether the NN performance can be improved still with a larger dataset. In Fig.~\ref{fig:scaling}, we picked an approximate metric for the NN performance to show its scaling with $n_{\rm sim,\, train}$, the number of simulations used for training. The approximate metric used here is simply the $\bar{N}_L^{\kappa\kappa}$, which is the average of $N_L^{\kappa\kappa}$ in the range $150 \leq L \leq 500$ (corresponding to the highlighted portion in grey on the blue NN curve in Fig.~\ref{fig:nlkk}). The rightmost point in Fig.~\ref{fig:scaling} with $n_{\rm sim}=19200$ corresponds to our fiducial choice, beyond which there seems to be diminishing return (the scaling is roughly linear in log space).

We shall see in the next section that though the lensing reconstruction performance seems to have slight room for improvement with training set size, the delensing performance with the current NN reconstructed $\kappa$ is nearly indistinguishable from the theoretical best solution provided by the converged iterative estimator.

\section{Delensing performance}
\label{sec:delensing_results}

 We now move on to compare the delensed performance of the various estimators by looking at the predicted delensed $B$-mode power spectrum $C_l^{BB,\,\rm delensed}$, obtained by averaging over a test set of 2400 observed CMB maps using the standard delensing pipeline as described in Sec.~\ref{sec:delensing}.

 \begin{figure}
    \centering
    \includegraphics[width=0.5\textwidth, keepaspectratio]{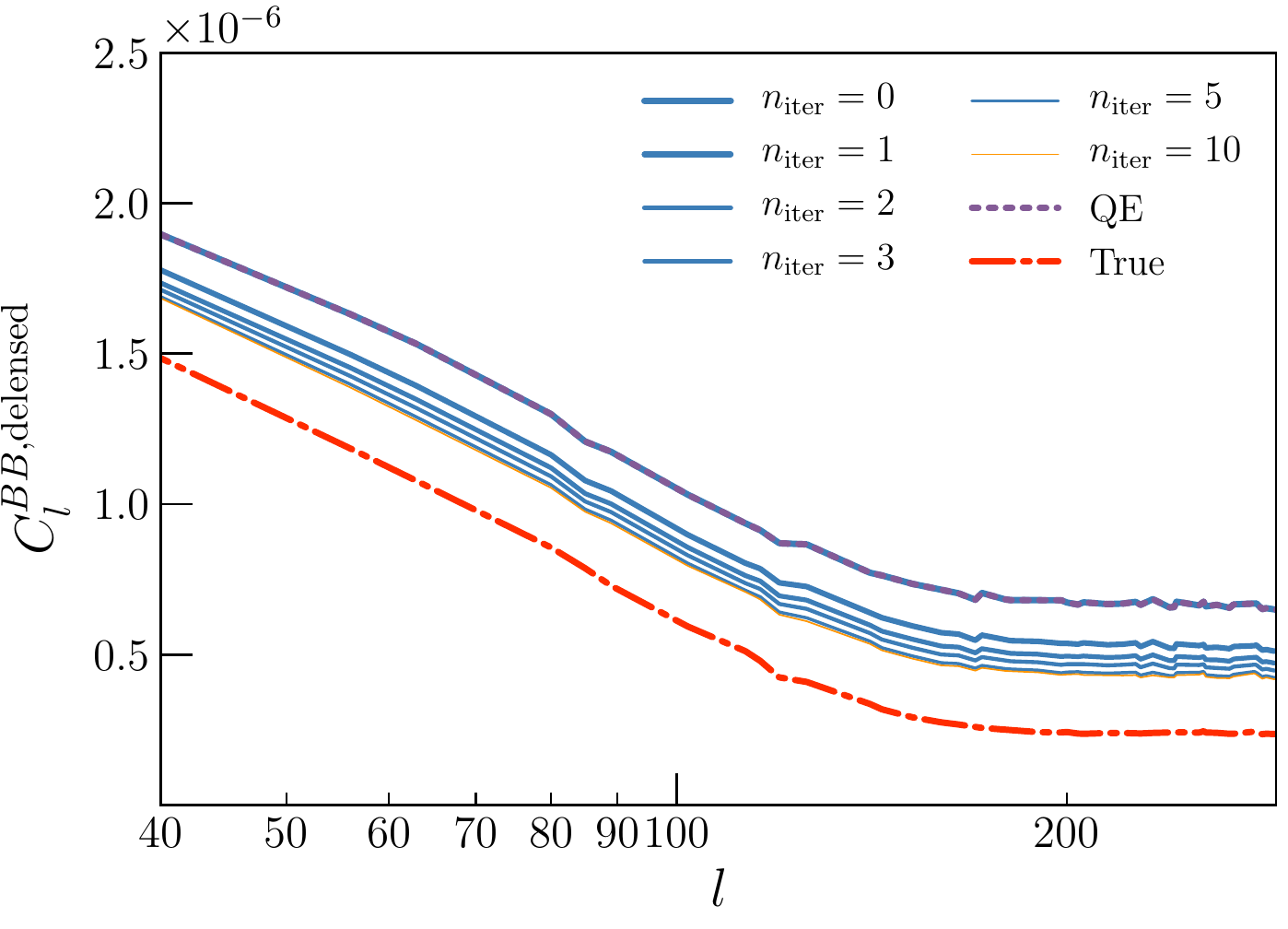}
    \caption{Noise-subtracted, delensed $B$-mode power $C_l^{BB, \, \rm delensed}$  calculated using the average of 2400 simulations, for various number of iterations $n_{\rm iter}$ of the iterative estimator (solid blue, decreasing thickness for increasing $n_{\rm iter}$). The $n_{\rm iter} = 10$ result (thin orange) is our fiducial result and is shown here to be converged. The quadratic estimator result (purple dashed) is the same as $n_{\rm iter} = 0$ in the iterative estimator as expected. We also show the delensed power using the true $\kappa$ map (red dash-dotted), which cannot be reached in reality because of the cosmic variance and other sources of error contributing to the overall reconstruction noise for $\kappa$.}
    \label{fig:clbb_delensed_vary_niter_r_0p01}
\end{figure}

\begin{figure}[h!]
    \centering
    \includegraphics[width=0.47\textwidth, keepaspectratio]{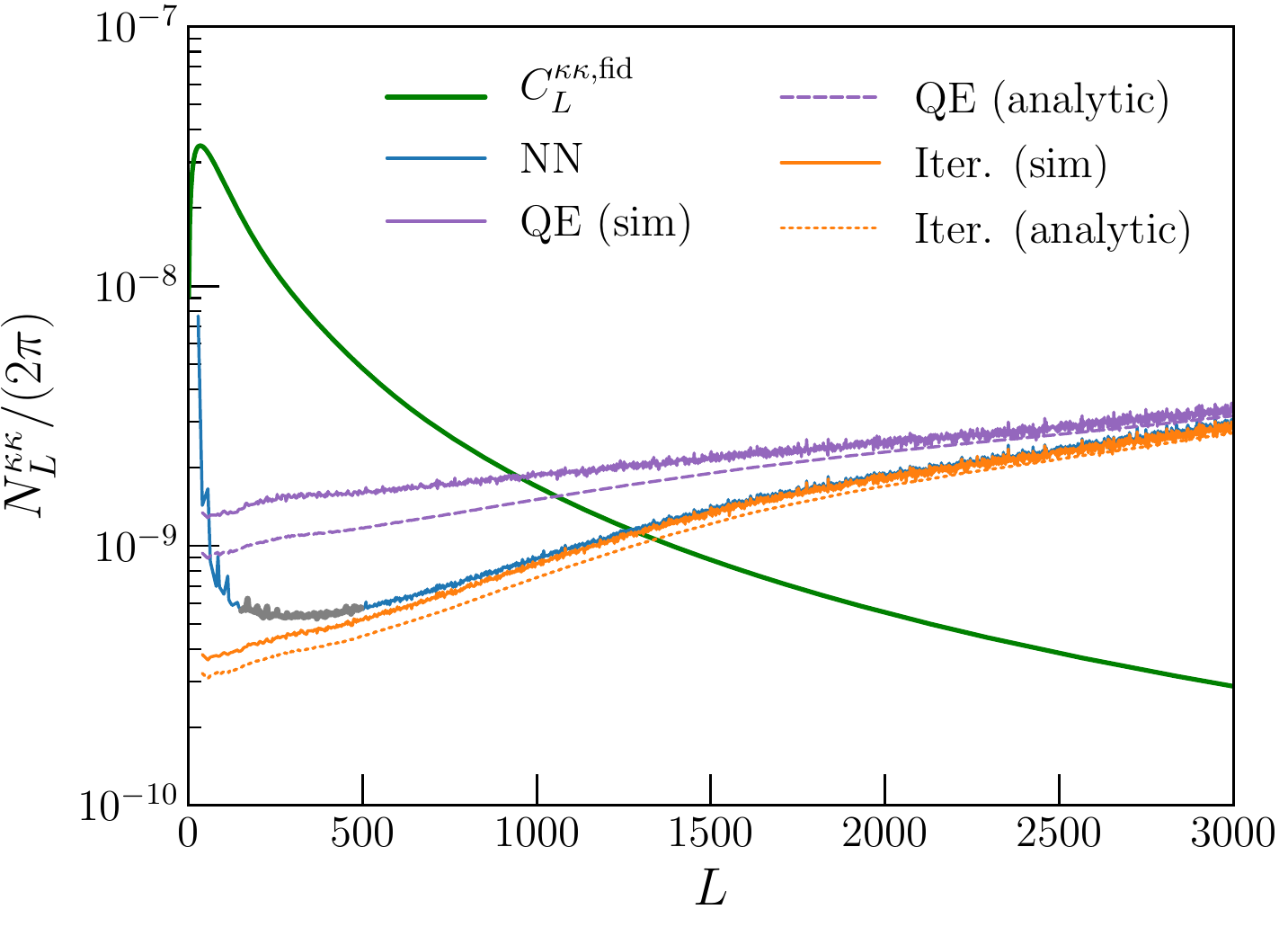}
    \caption{The analog of Fig.~\ref{fig:nlkk} but for the set of maps used for delensing, where the NN is now trained on $\lcut = 300$ CMB maps with a single output channel $\kappa$, and where all estimators are evaluated on $\lcut = 300$ CMB maps with $r = 0.01$ and default noise level. For an apples-to-apples comparison, we also added the QE and iterative estimator noise curves derived from simulations, similarly to what was done for the NN (see Eq.~\ref{eq:nlkk_NN}). The simulation-based noise curves are higher than the analytic predictions for the QE and iterative estimator, because they additional include the $N^{(1)}$ bias whereas the analytic curves only have $N^{(0)}$. The NN and iterative estimators noise curves are similar except for $L \lesssim 500$, but this does not affect the delensing performance as shown in Fig.~\ref{fig:clbb_delensed}.}
    \label{fig:nlkk_lmin_300_sim}
\end{figure}

\subsection{Quadratic and iterative estimator}

We calculate the quadratic estimator and iterative estimators on the test set of 2400 CMB observations using $\lensit$. Before estimating the lensing potential, the observed CMB maps are filtered to only contain $l \geq \lcut$. Otherwise, there would be spurious correlations between the observed $B$-mode on the scales we are interested in $l<\lcut$ and the $\kappa$ predictions since the latter would contain quadratic pairs constructed from $E$ and $B$-modes with $l<\lcut$~\cite{2011arXiv1102.5729T}.

We Wiener-filter the QE before using it in the delensing pipeline, where the Wiener filter is constructed using the analytic estimate of $N_L^{\kappa\kappa}$ for QE with $\lmin = 300$. For the iterative estimator, there is no Wiener-filter needed, since the converged estimator would already be the maximum a posteriori solution (at least exactly so for the case of $r = 0$). In Fig.~\ref{fig:kappa_maps}, we show an example of the $\kappa$ map predictions evaluated on the same set of observed $Q$ and $U$ maps for various estimators: The Wiener-filtered quadratic estimator (top right), the converged iterative estimator at 10 iterations (bottom left), as well as the true input $\kappa$ map (top left) which has more small-scale details.

In Fig.~\ref{fig:clbb_delensed_vary_niter_r_0p01}, we show the convergence of the $C_l^{BB,\, \rm delensed}$ for the iterative estimator as we vary $n_{\rm iter} = 0$ to 10.
Note that the $0^{\rm th}$ iteration is the same as the Wiener-filtered quadratic estimator since that is the starting point for the iterative estimator. The delensing result already converges for $n_{\rm iter} = 5$ (the thinnest blue line), but we will use $n_{\rm iter} = 10$ (orange line) as our fiducial result for the iterative estimator.

\begin{figure}
\centering
   \includegraphics[width=0.5\textwidth, keepaspectratio]{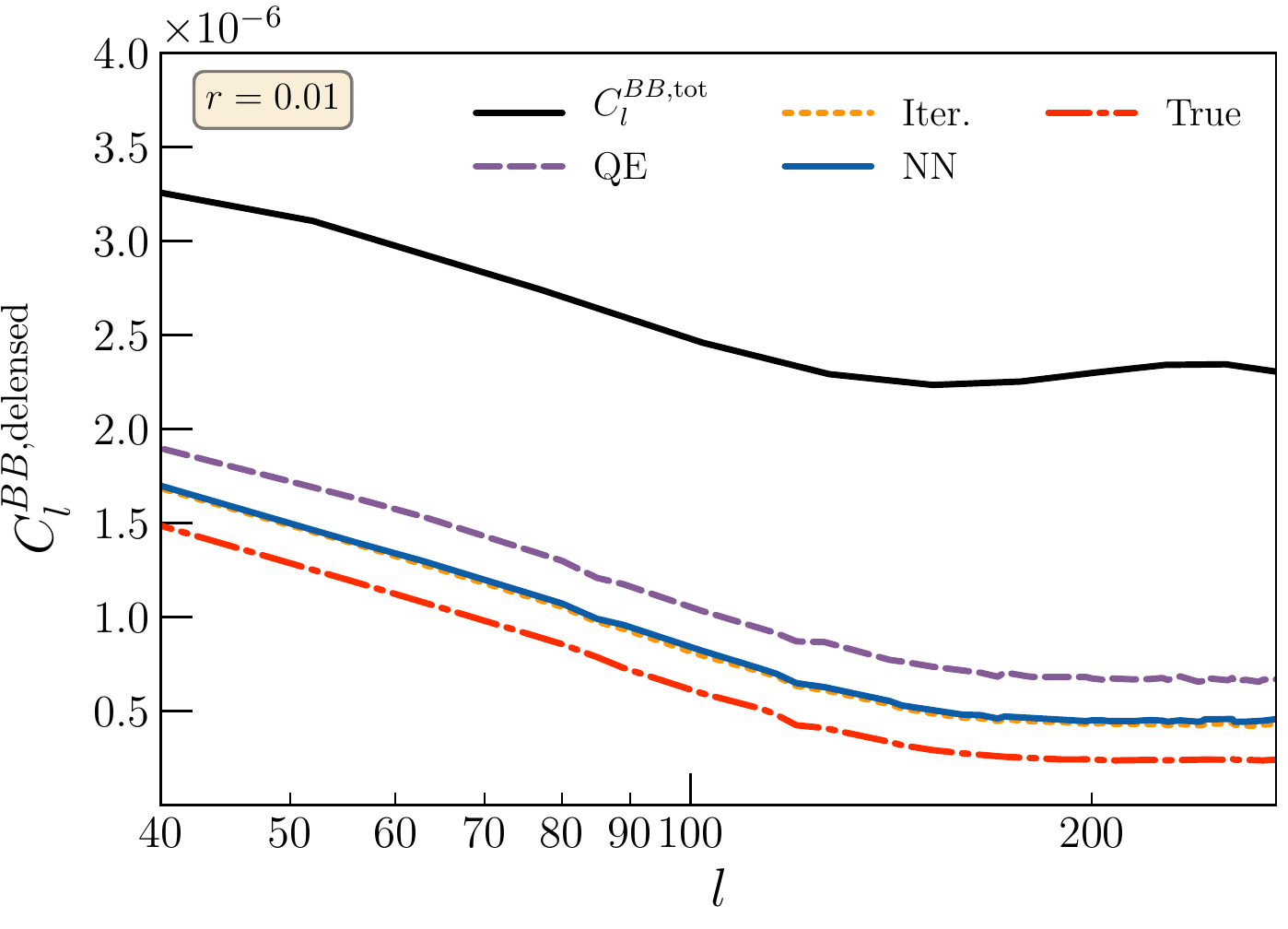}

   \includegraphics[width=0.5\textwidth, keepaspectratio]{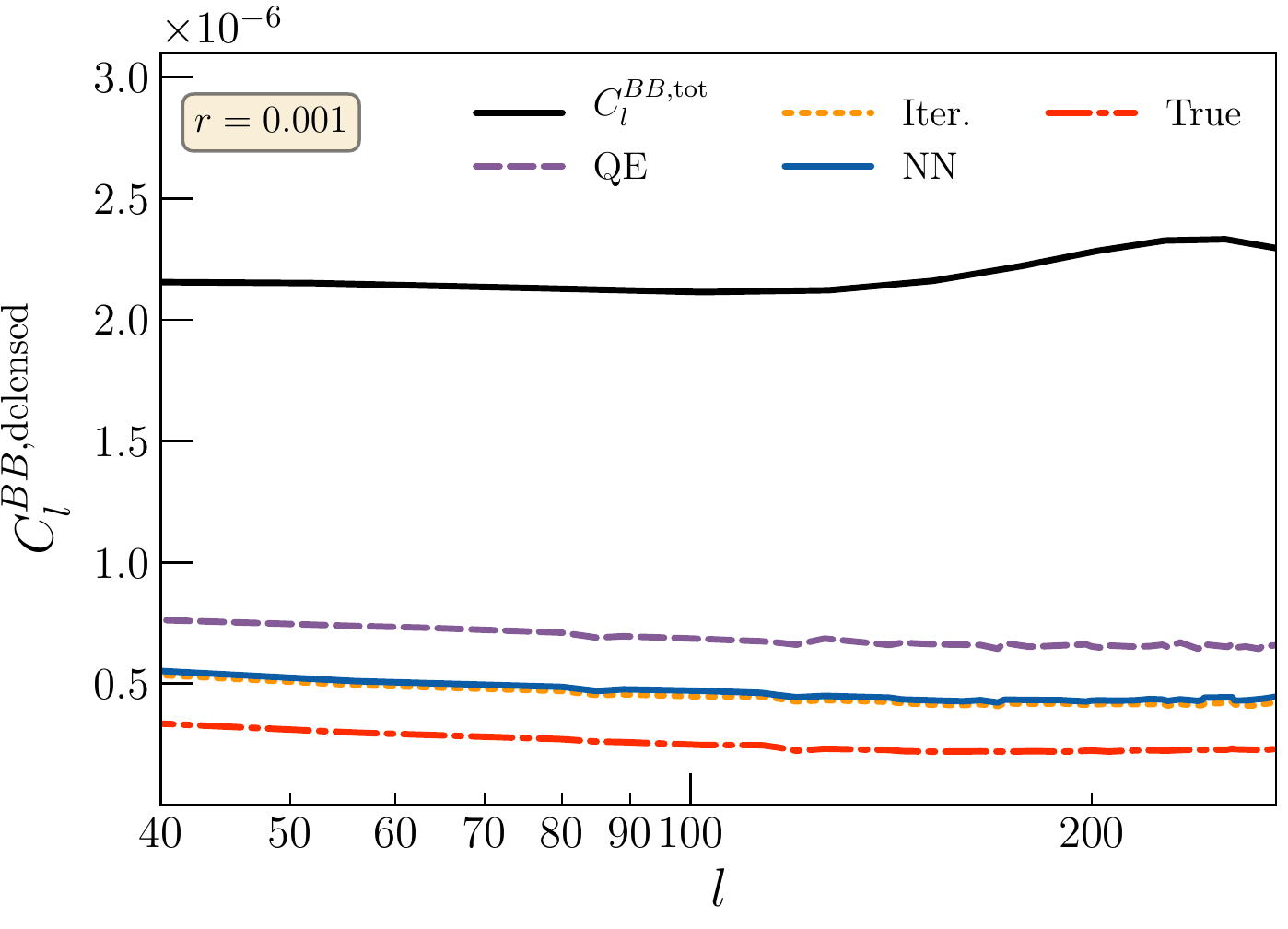}

\caption{Noise-subtracted, delensed power spectrum $C_l^{BB, \, \rm delensed}$ calculated using the average of 2400 simulations for various estimators: QE (purple dashed), iterative estimator (orange dotted) and NN (blue solid). We also plot for reference the noise-subtracted total $B$-mode power $C_l^{BB,\rm{tot}} = C_l^{BB, \rm lensed}+C_l^{BB, \rm tensor}$ (black solid) and the delensed power spectrum using the true $\kappa$ maps (red dash-dotted). }
\label{fig:clbb_delensed}

\end{figure}

\subsection{Neural network estimator}

For the delensing exercise, we will make use of the NN trained with only one output channel $\kappa$. Furthermore, it turns out that beside evaluating the NN model on a set of observed CMB maps with $l \geq \lcut$ as done in the case of the QE and iterative estimator, it is also necessary to apply the same cut to input $Q$ and $U$ maps during training. Otherwise the NN model ends up ``memorizing" information in the input $Q$ and $U$ maps at $l < \lcut$, and we end up with spurious correlations in the final delensed $B$-mode power spectrum. In other words, because the set of training maps sets an implicit prior for the NN model, there is information retained about $B$ at $l < \lcut$ in the trained model itself, and this information goes into the NN's $\kappa$ estimates, even when the trained model is only evaluated on observed $E$ and $B$ maps with $l \geq \lcut$.

We remind the reader that while the $E$ and $B$ maps used for evaluating the $\kappa$ estimators have $\lcut = 300$, it is not necessary to apply a cut to the Wiener-filtered observed $E$ maps used to construct the $B$-mode lensing template. The delensing performance would be worse if those modes were cut off, since they contribute to the $l<300$ lensed $B$-modes. Finally, to optimize the delensing performance, we Wiener-filter the NN predicted $\hat{\kappa}$ maps before feeding it into the delensing pipeline, where the Wiener filter is constructed using the noise power $N_L^{\kappa\kappa}$ estimated using the same test set. An example of the Wiener-filtered NN prediction for $\kappa$ is found on the bottom right of Fig.~\ref{fig:kappa_maps}. Features in the NN are similar to those in the converged iterative estimator, the optimal solution, and contains more details than the QE.

The delensed, noise-subtracted $B$-mode power spectrum for the NN is shown in blue in Fig.~\ref{fig:clbb_delensed}, where we have averaged over a test set of 2400 simulations with $r = 0.01$ and $r = 0.001$ in the top and bottom panels respectively.

\subsection{Comparison of all estimators}

In Fig.~\ref{fig:nlkk_lmin_300_sim}, we show a corresponding version of Fig.~\ref{fig:nlkk} but for the maps used for delensing, where the NN is trained on $\lcut = 300$ CMB maps with a single output channel $\kappa$, and where all estimators are evaluated on $\lcut = 300$ CMB maps. Instead of using the analytic predictions, we use the simulation-based noise curves, which are derived in the same way as for the NN (see Eq.~\ref{eq:nlkk_NN}). For the QE and iterative estimator, the simulation-based curves are higher compared to analytic predictions, mostly because they additionally include the $N^{(1)}$ bias. The $N^{(1)}$ bias might also be what explains the discrepancy seen in Fig. 9 of Ref.~\cite{Caldeira:2018ojb} for the $1\muKarcmin$ setup, between the analytic noise curve for the iterative estimator and the simulation-based NN noise curve. In our apples-to-apples comparison, the iterative estimator and the NN have very similar noise curves except for about $L \lesssim 500$. We will see next that this, however, does not affect the delensing performance.

We now compare all three estimators in Fig.~\ref{fig:clbb_delensed} in terms of their delensing performance for $r = 0.01$ and $r = 0.001$ in the top and bottom panels respectively, where we also include for reference the result from using the true $\kappa$ map while keeping the same Wiener-filtered observed $E$-modes as in all other estimators. Remarkably, the iterative and NN estimators have very similar performance.

This is the first demonstration that a NN-based estimator is capable of extracting all the information in the CMB maps \emph{needed for delensing}, at least in our idealized setup. The result builds confidence that the NN estimates of $\kappa$ are usable for delensing purposes, an important step beyond just comparing the two-point function of the estimator noise (as in Ref.~\cite{Caldeira:2018ojb} and in our previous section).

Compared to the iterative estimator, the NN method has the advantage that once the NN model is trained, its evaluation on data or simulations is almost instant. In our setup the NN took about 40 hours to train on the GPU including time to ensure its convergence, and 0.13 seconds to evaluate per realization, giving a total of about 5 minutes to evaluate all 2400 realizations. The iterative estimator on the other hand, takes no training time, but each realization takes about 1 minute to evaluate on a single CPU core for $n_{\rm iter}=5$ iterations, and about 2 minutes for $n_{\rm iter}=10$. So the total time of evaluating the delensing performance averaged over 2400 realizations was 40 hours (80 hours) for the iterative estimator with $n_{\rm iter}=5$ ($n_{\rm iter}=10$), similar to or more than the NN.

The difference in time would be even higher if more realizations were used, or if a larger sky area was observed. We also note that in the above calculation we did not count the training sample generation time for the NN, which is quite fast because of our simplistic setup, and given our data augmentation methodology which allows us to reduce the total number of independent simulations generated. This data generation step could of course become more involved if more realistic simulations are needed, by including for example, varying cosmology, varying models of foreground residuals, inhomogeneous and correlated noise, irregular masks, etc.


\section{Summary and discussion}
\label{sec:conclusion}

In conclusion, we used simulated observed CMB maps to predict the delensing performance of a neural network lensing estimator and compared it to the quadratic and iterative estimators.

To do so, we first trained the NN on a set of 19200 simulations where the input maps were the observed $Q$ and $U$ maps generated from a fiducial cosmology, varying values of $r$ and instrument noise, and where the target maps were the $\kappa$ and unlensed $E$ maps. We evaluated the trained model on a set of 2400 maps with fixed values of $r=0.01$ and noise level $1 \muKarcmin$ and found that the lensing reconstruction abilities of our NN are similar to those found in Ref.~\cite{Caldeira:2018ojb} with $r = 0$, whereas our $E$-mode reconstruction performed better (a similar conclusion was also found in Ref.~\cite{Guzman:2021nfk}). By looking at the scaling trend of the NN performance with the number of simulations used for training, it seems that it may be possible to further optimize the NN's performance in lensing reconstruction with many more simulations, though it would be in the regime of diminishing returns.

To assess the delensing performance, we used the standard delensing pipeline, where the $B$-mode templates are built by lensing the Wiener-filtered observed $E$-mode using the $\kappa$ estimator.
We found that for values of $r$ relevant for the next-generation CMB experiments, such as $r = 0.01$ and $r = 0.001$, the NN was able to attain the same optimal delensing performance achieved by the converged iterative estimator.
This was true despite the small discrepancy in the simulation-derived noise curves between the iterative estimator and the NN at $L \lesssim 500$.

We also found that it was necessary to train the NN on input CMB maps for which the scales of interest for the delensed $B$-modes are excised. This is similar to what is known for the quadratic and iterative estimators: One needs to evaluate these estimators on the observed CMB maps with $l \geq \lcut$, or there would be unwanted correlations between the delensed $B$-modes and the observed $B$-modes on scales $l<\lcut$. This is also true in the NN case, except that there is an additional step that the training must also be done on CMB maps with $l \geq \lcut$. The NN seems to be able to retain information about the observed CMB modes at $l < \lcut$ during training even when the trained model is only evaluated on maps with $l \geq \lcut$.

Moreover, we have used various techniques here in addition to what was previously explored in Ref.~\cite{Caldeira:2018ojb}. First, our training of the NN included observed CMB maps with tensor perturbations. The training set had a log-uniform distribution of $r$-values, so that the NN can learn from many examples of how the observed CMB maps change with $r$. We have also trained on uniformly distributed noise levels between [0, 2$\Delta^P_{\rm test}$], where $\Delta^P_{\rm test} = 1\, \muKarcmin$ is the desired polarization noise level used in the test set. Finally, we used data augmentation at each epoch during the training by randomly pairing pre-generated phases of scalar maps, tensor maps and noise maps, so that the NN has access in principle to $n_{\rm sim, train}^3 = 19200^3$ combinations if it was allowed to be trained for long enough.

In this work we have restricted ourselves to an idealistic setup, including cosmic variance for a fixed cosmology, varying tensor perturbations, homogeneous instrumental noise, no foregrounds and a square-shaped mask. In the future, it would be important to test how the NN does on more realistic setups such as varying cosmology, including foreground residuals, more complex instrumental noise models and irregular masks. The training techniques we used here, which incorporate a distribution of tensor and noise models, can be further extended to train the NN with a range of cosmology and foreground contamination models all at once, as well as other models that have significant theoretical uncertainties in them. One may also explore varying cosmological parameters with a similar setup, or even predicting them as auxiliary output of the NN to help guide the training (e.g. \cite{Pan:2019vky}).

Once fully trained, the advantage of the NN is that it is very fast to evaluate, so studying its noise properties and delensing performance using a large set of simulations would be faster than for the iterative estimator, which has a much longer evaluation time per realization. So although processing the data once is faster for the iterative estimator than for the NN (with training time included), the total compute time needed for validating the analysis pipeline on a large set of simulations could be smaller for the NN (unless the generation of training data set starts to take up a significant amount of time).

The iterative estimator's performance in the presence of realistic foreground residuals -- whether they would degrade the optimality or introduce any bias -- is yet to be tested. It would be interesting to determine if the NN could perform better compared to the iterative estimator under certain scenarios. More futuristically, techniques are being developed to infer foreground contaminations in CMB observations using NNs (e.g.~\cite{Farsian:2020adf, Petroff:2020fbf}) and may be combined into a multi-stage or hybrid NN to remove the foreground contaminations at the same time as the lensing reconstruction.

Finally, given the ability of the NNs to model nonlinear mappings, it would be interesting to see whether they could perform better in reconstructing the lensing convergence where other nonlinear effects are included, for example, the post-Born effects (e.g.~\cite{Pratten:2016dsm}) which account for the interaction between more than one lensing deflections along the line-of-sight, or effects from the nonlinear power spectrum.

In conclusion, we used a neural-network-based lensing reconstruction in a standard delensing pipeline, demonstrating optimal delensing performance, achievable so far only by the iterative estimator, on a simplified setup in the range of $r$ relevant to the next-generation experiments ($r\lesssim 0.01$) and for CMB-S4-like noise level. This study paves the way for using deep learning in delensing analyses, and provides an alternative to the iterative estimator. The method represents a successful hybrid of deep-learning and standard cosmology techniques, making use of both our prior knowledge of physics as well as computational advances in deep learning.


\begin{acknowledgements}

Chen H.: This paper is dedicated to my Maker, who has provided the ability to do this scientific work, guidance in times of difficulties, all the resources to conduct the work, and the precious people who shaped the experience in invaluable ways -- Trey Driskell and Chris Heinrich. Chen H. would also like to thank Olivier Doré for early stage feedback on the work and for co-authoring the following grant which partially supported the work: Data Science Pilot Grant at Jet Propulsion Laboratory. We are especially grateful to Julien Carron for providing guidance on LensIt and extensive feedback on the manuscript, and also thank Olivier Doré and Wayne Hu for their very useful discussions and feedback on the manuscript. We are also grateful to the Texas Advanced Computing Center (TACC) for providing the computing resources needed in this work.

\end{acknowledgements}

\bibliography{lensnet.bib}

\appendix

\end{document}